%% file: main.tex
\documentclass[sigconf]{acmart}
\AtBeginDocument{%
  }
\input{tables}
\usepackage{multirow}
\setcopyright{acmlicensed}
\copyrightyear{2026}
\acmYear{2026}

\usepackage[breakable, skins]{tcolorbox}
\usepackage[dvipsnames]{xcolor}
\usepackage{soul}
\usepackage{algorithm}
\usepackage{algpseudocode}
\tcbuselibrary{skins}
\newenvironment{summarybox}
{\begin{tcolorbox}
[breakable,enhanced,arc=0mm,colback=gray!10,frame hidden,overlay broken={%
    \draw[thick,black] (interior.north west)--(interior.south west);
},overlay unbroken={%
    \draw[thick,black] (interior.north west)--(interior.south west);
},left=2pt,right=0pt,top=0pt,bottom=0pt,before={\vspace{3pt}\noindent},after={\vspace{0pt}}]
\setlength{\baselineskip}{0.75\baselineskip}}
{\end{tcolorbox}}
\newenvironment{summary}
{\vspace{5pt}\noindent\begin{summarybox}}
{\end{summarybox}\vspace{-5pt}}

\begin{document}
\renewcommand{\hl}[1]{#1}
\title{Towards Green AI: Decoding the Energy of LLM Inference in Software Development}

\author{Lola Solovyeva}
\email{o.solovyeva@utwente.nl}
\orcid{0009-0008-6903-7086}
\affiliation{%
  \institution{University of Twente}
  \city{Enschede}
  \country{the Netherlands}
}

\author{Fernando Castor}
\email{f.castor@utwente.nl}
\affiliation{%
  \institution{University of Twente}
  \city{Enschede}
  \country{the Netherlands}
}


\begin{abstract}
\textbf{Context:} AI-assisted tools are increasingly integrated into software development workflows, but their reliance on large language models (LLMs) introduces substantial computational and energy costs. Understanding and reducing the energy footprint of LLM inference is therefore essential for sustainable software development.
\textbf{Objective:} In this study, we conduct a phase-level analysis of LLM inference energy consumption, distinguishing between the (1) \textit{prefill}, where the model processes the input and builds internal representations, and (2) \textit{decoding}, where output tokens are generated using the stored state. 
\textbf{Method: } We investigate six 6B-7B and four 3B-4B transformer-based models, evaluating them on code-centric benchmarks HumanEval for code generation and LongBench for code understanding.
\textbf{Results: } Our findings show that, within both parameter groups, models exhibit distinct energy patterns across phases. Furthermore, we observed that increases in prefill cost amplify the energy cost per token during decoding, with amplifications ranging from 1.3\% to 51.8\% depending on the model. Lastly, three out of ten models demonstrate \textit{babbling} behavior, adding excessive content to the output that unnecessarily inflates energy consumption. We implemented \textit{babbling suppression} for code generation, achieving energy savings ranging from 44\% to 89\% without affecting generation accuracy.
\textbf{Conclusion: } These findings show that prefill costs influence decoding, which dominates energy consumption, and that \textit{babbling suppression} can yield up to 89\% energy savings. Reducing inference energy therefore requires both mitigating \textit{babbling} behavior and limiting prefill’s impact on decoding.
\end{abstract}

\maketitle

\section{Introduction}
\par  AI-assisted tools are increasingly integrated into software development processes~\cite{GenAISoftware, 10.1145/3708525, 10.1145/3643795.3648379, 10.1145/3696630.3730563}.  Those who adopt them believe they boost productivity, reduce costs, and offer valuable coding assistance~\cite{10628428}.  Two widely used examples of these tools, GitHub Copilot~\cite{githubCopilot} and ChatGPT~\cite{chatGPT}, have been the focus of studies examining both their effectiveness in code generation and the level of trust users place in them~\cite{10.1145/3639477.3643648}. Both rely on large language models (LLMs), which are typically based on the transformer architecture and are trained on massive amounts of text and code to generate its output.
\par While they can accelerate the process of building software, their inference processes introduce a non-trivial energy cost, particularly when used repeatedly in CI/CD pipelines or large-scale maintenance workflows~\cite{10.1145/3708525}. For example, ChatGPT was queried approximately 13 million times in January 2023, and the total energy consumed during inference \hl{in that month alone} already exceeded that of its training~\cite{10549890}. \hl{Research firm SemiAnalysis} suggested that OpenAI required 3,617 of NVIDIA’s HGX A100 servers, with a tottal of 28,936 GPUs , to support ChatGPT, implying that it requires 564 MWh per day for its inference~\cite{SemiAnalysis, DEVRIES20232191}. Meanwhile, an estimate of 1,287 MWh was used in the training phase of  GPT-3. Furthermore, model training can be scheduled during periods when a larger share of energy comes from low-carbon or renewable sources, thereby reducing the $\text{CO}_2$ emissions associated with training. In contrast, inference is typically time-sensitive and must be executed immediately, leaving little flexibility for such optimization. As a result, the overall sustainability of the software lifecycle now also depends on the efficiency of the AI tools that support them.
\par While existing studies on LLM efficiency focus on architectural techniques, these approaches often treat inference as a uniform process~\cite{10549890,10.1145/3757892.3757900, jenga}. In practice, inference consists of two distinct phases: prefill and decoding~\cite{10.5555/3691938.3691945, fernandez-etal-2025-energy, 11113611}. The \textbf{prefill phase} phase occurs when the model first processes the input prompt to build internal key–value caches for attention. The \textbf{decoding phase} follows, generating output tokens autoregressively for each request, using the cached representations from the prefill step. From a resource utilization perspective, LLM inference is challenging, as its phases place varying demands on different resources at different times. The prefill phase is highly parallel and therefore compute-bound, whereas the autoregressive nature of the decoding phase makes it more memory-bound~\cite{10.5555/3691938.3691945}. \hl{Treating LLM inference as a uniform process therefore risks overlooking these phase-specific characteristics, potentially obscuring opportunities to optimize for energy efficiency.}
\par The workload for both phases depends on the size of the input and the length of the output. \hl{The nature of the task an LLM is used to perform influences the input size and the expected output length.} \hl{In software development, function-level code generation typically requires relatively short inputs, often consisting only of a docstring and a function signature, while the generated output may exceed the input length depending on the complexity of the function. Therefore, longer outputs place greater emphasis on the decoding phase, making it a critical target for optimization to achieve energy savings.} On the other hand, \hl{long-context tasks, such as code repository understanding, involves processing larger inputs, but generating shorter output, thereby placing more emphasis on the efficiency of the prefill phase.} Hence, it is clear that the choice of optimization strategy should depend on the specific task the model is intended to perform.
\par In this work, \hl{we demonstrate that ten decoder-only transformer models exhibit different energy consumption patterns across the prefill and decoding phases, despite having comparable number of parameters.} We conduct empirical experiments on models from various families and versions (e.g., Llama, Phi, Gemma, Qwen) with different parameter sizes (3B-4B, and 6B-7B), performing software development tasks, \hl{function-level code generation and code repository understanding}, to evaluate their energy consumption during both phases. Hence, our study has the following contributions: 
\begin{enumerate}
    \item Phase-level analysis of LLM inference, revealing how both phases exhibit distinct energy consumption characteristics.
    \item Empirical experiments on 10 transformer models (Llama, Phi, Gemma, Qwen) of comparable parameter sizes (3B-4B, 6B-7B), using software development tasks (code generation and code understanding) to quantify phase-level energy patterns for different workloads.
    \item \hl{Analysis of the impact of prefill costs on decoding efficiency, demonstrating how higher prefill overhead can amplify per-token energy consumption during decoding.}
    \item \hl{Identification of energy inefficiencies caused by babbling behavior, where excessively verbose outputs lead to unnecessarily high energy consumption.}
\end{enumerate}
\par \hl{Our findings show that decoder-only transformer models with comparable parameter counts can nevertheless exhibit substantially different phase-level energy consumption patterns, suggesting that low-level implementation choices (e.g., operator scheduling, memory management and allocation) play a significant role in determining energy efficiency. } We reveal how energy costs evolve during decoding, showing that the energy per token grows as new tokens are generated, where the speed of growth depends on the model. Furthermore, we show that input size affects both phases, increasing prefill costs and the initial token cost in decoding, highlighting the influence of workload characteristics on energy consumption. Lastly, we identify that some models \textit{babble}, generating unnecessary output beyond task completion, increasing energy usage. The replication package for this study is publicly available~\cite{phaseLevelEnergyAnalysis}.



\section{Related work}
\par With the widespread adoption of LLMs in software development, there has been a growing body of research focusing on their evaluation, including functional correctness~\cite{alizadeh2025languagemodelssoftwaredevelopment, humanevalPlus, hallucinationsCodeGen} and output quality~\cite{ai-powered-power-hungry, codeQualityLLM, apsan2025, saad2025senaisoftwareengineeringnative}. Recently, as the carbon footprint of these models has become a growing concern, attention has also shifted toward their environmental sustainability~\cite{wattsLuccioni,greenreviewLLMs, 11305123}. At present, the overall sustainability of the software development lifecycle is \hl{partially} dependent on the efficiency of the AI tools that support it.
\par Some works~\cite{ai-powered-power-hungry, apsan2025} have focused on evaluating the energy efficiency of LLM-generated code, motivated by the idea that code produced by these models should not be more energy-consuming than human-written code, considering its deployment at scale could result in substantial increases in energy consumption and overall computational cost. Solovyeva et al.~\cite{ai-powered-power-hungry} showed that \hl{energy efficiency of LLM-generated code is comparable to human-written code and, in some cases, can even be more energy-efficient,} whereas the opposite trend holds for Java and C++. The findings of Apsan et al.~\cite{apsan2025} complement these results, showing that LLM-generated Python code demonstrates higher energy efficiency than human-written code, but only when executed on personal computers. They observed that this advantage disappears in server and resource-constrained environments. While their work focused on evaluating the energy efficiency of the code generated by LLMs, our work focuses on analyzing the energy consumption of the models themselves during the inference.
\par \hl{Previously, environmental concerns regarding LLMs focused primarily on the training phase, due to its heavy computational requirements and long training times}~\cite{Verdecchia2023ASR}. \hl{However, with the widespread deployment of LLMs and their increasing usage, the inference phase has also become a significant concern}~\cite{DEVRIES20232191},\hl{ leading to a growing number of studies analyzing its energy requirements.}
Luccioni et al.~\cite{wattsLuccioni} conducted a comprehensive study comparing the inference energy costs of various generative models, covering tasks from text classification to image generation. They show that energy consumption is influenced by the model’s architecture, its modality, and the nature of the task. Jegham et al.~\cite{jegham2025hungryai} compared the inference energy consumption of 30 state-of-the-art LLMs, confirming that architectural differences significantly influence energy usage. They concluded that although AI systems are becoming faster and more cost-effective, their widespread adoption leads to disproportionate increases in overall resource consumption. In the context of software development, Alizadeh et al.~\cite{alizadeh2025languagemodelssoftwaredevelopment} analyzed accuracy–energy trade-offs across multiple software development tasks and showed that energy consumption varies substantially by task and model architecture. Complementing this, Mehditabar et al.~\cite{mehditabar2025smartcostlybenchmarkingllms} introduced BRACE, a unified benchmark that evaluates coding LLMs by jointly considering functional correctness and energy efficiency to support sustainability-aware model selection.
Unlike prior studies that treat LLM inference as a single process, our work examines its energy efficiency across the prefill and decoding phases, narrowing the gap in understanding how architectural design influences overall inference energy consumption.
\par Several studies have explored reducing LLM inference energy consumption through system-level optimizations. Niu et al.~\cite{10.1145/3757892.3757900} conducted a comprehensive evaluation of energy efficiency across several widely used LLM inference engines, including vLLM, DeepSpeed, TensorRT-LLM, and Transformers. Their analysis focused on two stages: engine initialization with model loading, and the token generation stage. In contrast, our study breaks down the token generation stage further, examining the prefill and decoding phases separately to better understand their individual contributions to overall energy consumption. Zhang et al.~\cite{jenga} developed a memory allocation framework for heterogeneous LLMs, focusing on improving GPU memory utilization. The authors do not consider the energy usage of the GPU in their work. Furthermore, GPU memory utilization can differ between the prefill and decoding phases due to their distinct computational and storage requirements. Maliakel et al.~\cite{maliakel2025} and Stojkovic et al.~\cite{stojkovic2024} investigated adaptive resource allocation and GPU frequency tuning to accommodate varying batch sizes, input lengths, and output lengths, showing how these factors influence the energy consumption of LLM inference. Our study complements theirs, as the prefill and decoding phases have distinct computational and memory requirements, suggesting that resource allocation and GPU frequency mechanisms should consider these phase-specific differences. 
\par Lastly, a few prior \hl{papers} have examined the energy consumption of the prefill and decoding phases individually. Fernandez et al.~\cite{fernandez-etal-2025-energy} conducted an empirical analysis covering software frameworks, decoding strategies, GPU architectures, and model parallelism configurations. \hl{The analysis of energy consumption during the prefill and decoding phases was not the primary focus of their work, however they demonstrated the effect of sequence length on both phases. Extending their findings, we demonstrate the impact of the prefill phase on decoding and provide a more fine-grained analysis of energy consumption during the decoding phase.} Fan et al.~\cite{11113611} developed an energy-efficient LLM inference system for edge heterogeneous platforms by splitting the prefill and decode phases across devices and dynamically selecting the optimal execution plan based on model, hardware, and input characteristics. \hl{However, their goal is system design, whereas our goal is to identify energy patterns in both phases, their interactions, and associated inefficiencies, as well as to understand whether these patterns hold across different model families. }


\section{Methodology}
\par By utilizing the formulation proposed by Basili et al.~\cite{Basili}, the high-level goal of this study is to \textit{analyse} prefill and decoding phases of LLM inference \textit{for the purpose of} evaluation \textit{with respect to their} energy consumption and contribution to the inference  \textit{from the view point of} software engineers \textit{in the context of} code generation and code understanding. This can be summarized in the following primary research question:
 \begin{quote}
        \emph{RQ: \hl{How do the two phases contribute} to the total energy consumption of LLM inference for code generation and code understanding?}
    \end{quote}
To answer the main research question, we measure the energy consumption in each phase of LLM inference and assess how much each phase contributes to the total energy consumption of the inference. To gain a deeper and more comprehensive understanding of the topic, the primary research question is further divided into the following sub-questions:\\
\textbf{\underline{RQ1}: }\textit{To what extent do the magnitudes of input and output workloads affect the contributions of both phases to the overall energy consumption of LLM inference?} 
\\
\textbf{Motivation.} The intensity of an LLM inference workload is determined by the size of the input and the length of the output. In the prefill phase, workload intensity increases with longer inputs, since all input tokens must be processed. In contrast, the decoding phase is influenced primarily by output length, as it generates the remaining tokens sequentially. Because the prefill phase processes tokens in parallel while the decoding phase generates tokens sequentially, increasing the workload can affect the energy contribution of each phase differently. \hl{Answering this question helps determine the primary phase to optimize based on the workload characteristics.} \\
\textbf{\underline{RQ2}: }\textit{To what extent do these findings generalize across models of different sizes and families?} 
\\
\textbf{Motivation.} Models from different families can differ in intrinsic characteristics. At a lower level, kernel implementations, such as matrix multiplications, attention, and activation functions can vary in precision, memory layout, fusion, and parallelization strategy. These variations, along with differences in numerical precision, software versions, and hardware optimizations, affect the models’ outputs and performance. \hl{Therefore, it is important to consider a wide spectrum of models to determine whether the same energy patterns hold across both phases. Answering this question also helps clarify whether model implementations play a significant role in influencing energy consumption during the prefill and decoding phases.}
\subsection{Variables}
\par This study involves the following independent variables:
\begin{enumerate}
    \item \textbf{LLM.} We categorize the evaluated LLMs into two groups: the 6–7B group and the 3–4B group. The models were selected based on their \hl{accuracy score} on the HumanEval dataset, as reported in the \textit{Big Code Models Leaderboard}~\cite{BigCodeLeaderboard}, and according to the hardware constraints of our setup, since the models had to run locally. Additionally, the models within each group were chosen to have a comparable number of parameters, ensuring a fair comparison.
    \item \textbf{Workload.} To evaluate the contribution of both phases, we need to examine how they respond to increases in input size and output length. However, constraining either variable to a specific value is challenging and unrealistic: input length depends on the task and dataset, while output length can be limited but the model may terminate earlier than the maximum. Hence, we design five workloads based on the task and the approach used to solve it. The benchmarks and workloads are described in detail in the Section~\ref{subsec:workloadsBench}.
\end{enumerate}
\par For the dependent variables, we have the following list: 
\begin{enumerate}
    \item \hl{\textbf{GPU energy (J)}}: \hl{We focus on GPU energy, since it dominates energy consumption of LLM inference. We record it per single inference, averaged over a given workload.}
    \item \textbf{Prefill energy (J)}: Energy consumption specifically during the prefill phase, referring to the energy required to process the input before generating the first output token.
    \item \textbf{Energy per token (J)}: Total inference energy divided by the number of generated tokens, representing the average energy cost per token, which is a common metric to report in similar studies. This metric does not account for phase separation, thus it includes the energy costs of prefill and decoding. 
    \item \textbf{Energy per token in decoding (J)}: Since we measure energy on a per-token basis, we sum the energy consumed by each generated token after the prefill phase and divide it by the number of tokens produced during decoding. This metric does not include the cost of prefill. Hence, it helps to evaluate the costs of decoding separately. 
    \item \textbf{Accuracy (\%)}: For code generation, we use the \textit{pass@1}, which measures the proportion of generated solutions that successfully pass the benchmark test cases. In the code understanding task, the LLM addresses multiple-choice questions, and accuracy is determined by the number of correct answers. The responses are manually analyzed.
    \item \textbf{Output}: The number of tokens generated is also recorded. Although a maximum token limit may be imposed, the model can terminate earlier upon generating the end-of-sequence token.
\end{enumerate}

\subsection{Workloads \& Benchmarks}
\label{subsec:workloadsBench}
\par In this study we consider three different settings for code generation, using \textbf{HumanEval}~\cite{humanEval} as our benchmark:
\begin{enumerate}
    \item \textbf{Zero-shot prompting (\texttt{0-shot})}: In this context, instruction, docstring, and function's signature are provided. The input is intentionally minimal, whereas the output is required to be a fully generated function. The input varies between 50 and 400 tokens, depending on the complexity of the prompt. The maximum output length is set to 300 tokens, since around 90\% of the canonical solutions fall under 450 characters. \hl{A single token can represent a portion of a word and often contains multiple characters. Hence, the choice of 300 tokens is sufficient to cover 450 characters. This setting serves as a baseline, with both input and output lengths kept relatively short and comparable. } 
    \item \textbf{Few-shot prompting (\texttt{2-shot})}: In this setting, two illustrative examples are provided to guide the model’s response prior to giving the instruction and function signature. Therefore, the input length is about three times greater than in the zero-shot scenario, while the expected output remains unchanged from the previous case. Hence, the variation in input length changes from previous workload to the range between 150 and 1200 tokens, whilst the maximum allowed output is still set to 300 tokens. \hl{This setting deviates from the baseline by using a longer input. We chose \texttt{2-shot} prompting as it increases input length more substantially than \texttt{1-shot}.} 
    \item \textbf{Zero-shot chain-of-thought (\texttt{0-shot CoT}):} \hl{This setting deviates from the baseline by increasing the output length.} Hence, it replicates the zero-shot prompting setup, except that the model is explicitly instructed to reason step by step prior to output generation. Hence, the input length is slightly increased from the \texttt{0-shot} workload and ranges between 60 and 410 tokens. However, the resulting output is longer than in previous workloads, containing both the reasoning process and the final generated function. We limited the models to 2000 tokens, corresponding to over 4000 characters, because the functions are not highly complex and would not require extensive reasoning. 
\end{enumerate}
\par Furthermore, we employ \textbf{LongBench}~\cite{longBench}, a benchmark specifically designed to evaluate how LLMs handle long-context scenarios, with input lengths ranging from 8K to 2M words. It includes a code repository understanding task, in which the user provides both a question and a relevant portion of a code repository for the model to reference when generating an answer. Hence, this setup allows us to assess model performance in situations where the input is substantially longer than the output. We consider two settings for this benchmark:
\begin{enumerate}
    \item \textbf{Code understanding (\texttt{CU})}: In this workload, the input ranges between 4000 and 8000 tokens. The maximum number of input tokens was set based on the available GPU memory. Since the dataset consists of multiple-choice questions, the model’s output is expected to be a single letter. Hence, the maximum output length was set to 10 tokens.
    \item \textbf{Code understanding with explanation (\texttt{CU-long}):} The input size stays the same as in the previous workload, but the maximum output length is extended to 300 tokens, allowing the models to reason beforehand and provide an explanation for the answer. 
\end{enumerate}
\subsection{Design}
\par To address the research question, we use empirical data collected from our experiments, in which 10 LLMs were evaluated across 5 workloads, resulting in 50 experimental trials. \hl{Each trial was executed once. The trial duration varies depending on the workload and model, ranging from a minimum of 5 minutes to a maximum of 171 minutes.} The same procedure was followed for each workload. First, the model and tokenizer were loaded onto the GPU. Next, we applied the methodology proposed by Babakol et al.~\cite{methodology_babakol}, where two parallel processes are executed for measurement: (1) \hl{GPU} energy consumption is sampled every \hl{0.1} seconds with corresponding timestamps, and (2) LLM inference is performed with timestamps recorded for each generated token at the end of its generation. \hl{The timestamp for the start of the first token generation is recorded immediately after the input prompt is passed to the LLM, capturing the beginning of the prefill phase.} The raw measurements are recorded for every inference run. HumanEval includes 164 data entries, while LongBench contains 50 data entries. This results in a corresponding number of inferences per model, depending on the workload. 
\tableResultsModels
\begin{figure}[h!]
    \centering
    \includegraphics[width=0.6\linewidth]{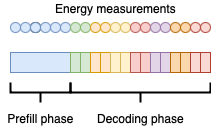}
    \caption{Simplified methodology proposed by Babakol et al.~\cite{methodology_babakol}, aligning token generation and energy measurements based on timestamps. \hl{Each color represents a different token.}}
    \label{fig:methodology}
\end{figure}
\par \par The measurements are then aligned based on their timestamps, as illustrated in Figure~\ref{fig:methodology}. \hl{The energy samples for token \textit{n} are collected between the timestamp of the generation of token \textit{n‑1} (which represents the end of its generation) and the timestamp for token \textit{n}. These samples are then aggregated to represent the total energy for token \textit{n}.} Based on this per-token energy aggregation, we can determine the energy consumed during the prefill phase, which corresponds to the generation of the first token, and the energy used during the decoding phase, calculated as the sum of the energy spent on all subsequent tokens. 
\par \textbf{Evaluation.} We use provided HumanEval test cases to assess the correctness of generated solutions. For all settings of code generation we employ pass@1, that shows the probability that the model generates a correct solution on its first attempt. For code understanding, LongBench includes ground-truth answers, which are used to compare against the model-generated responses. In this case, accuracy is defined as the proportion of answers that match the ground truth.

\subsection{Experimental setup}
\subsubsection{\textbf{Measurement environment}}
The experiments were conducted using UT-JupyterLab 3.6.8~\cite{UTJupyterLab}  with IPython 8.21.0 and Python 3.10.12. It provides access to a GPU cluster, where we selected NVIDIA A10 GPU with 24GB of memory.
All experiments were run on a single machine, ensuring exclusive access with no other processes running. Before each inference run, the system was checked for active processes, and inference was executed only when the machine was confirmed to be free. To record GPU power consumption, we used pyNVML~\cite{pynvml}, a Python binding of the NVIDIA Management Library~\cite{nvidia:2025}. The sampling frequency was set to 10Hz, as we measure energy consumption per token during inference, which requires high-precision measurements. The energy is then determined as a product of average power and time. 
\subsubsection{\textbf{LLMs and hyperparameters}}The LLMs were loaded using HuggingFace \texttt{transformers} 4.56.1 and executed using PyTorch 2.8.0. For all experiments, the top-p value was set to 0.95 and the temperature to 0.1, ensuring a fair comparison between the models. Because our evaluation relies on the pass@1 metric, we set a low temperature to favor the selection of the most probable tokens and ensure more deterministic outputs. Lastly, all the prompts followed the same template depending on the chosen workload, ensuring that every model received identical input. 

\section{Results}
\begin{figure}[t!]
    \centering
    \includegraphics[width=0.8\linewidth]{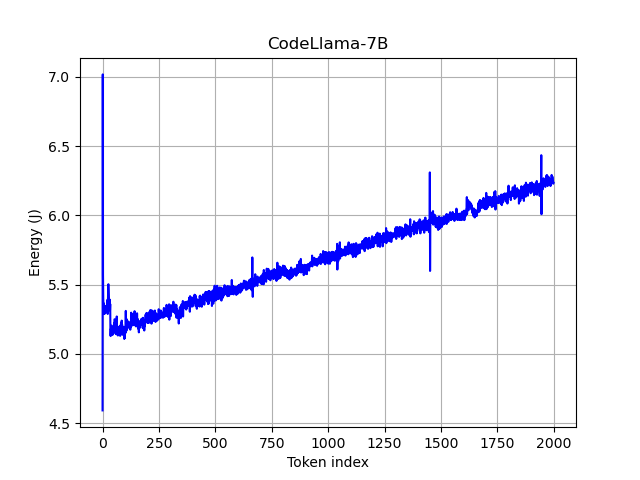}
    \caption{Energy consumption per token during token generation for \texttt{CodeLlama-7B} for \texttt{0-shot CoT}. The x-axis shows token index and y-axis energy consumption for that token. }
    \label{fig:codeLlamaTokenGen}
\end{figure}
\par Table~\ref{tab:resultsTotal} presents the results for the accuracy of the models across all workloads and benchmarks. For code generation, the highest overall accuracy of 74\% was achieved by \texttt{Qwen3-4B} under the \texttt{0-shot CoT} workload. However, \texttt{Qwen2.5-Coder-7B} consistently demonstrates the highest or among the highest \hl{accuracy} across all workloads. Two smaller models, \texttt{Phi4-4B} and \texttt{Qwen2.5-Coder-3B}, achieve nearly comparable performance despite their smaller size.
\par Regarding code understanding, \texttt{Qwen2.5-Coder-7B} achieved the highest accuracy of 36\% after being instructed to reason about the answer. Overall, most models, except for \texttt{CodeQwen1.5-7B} and \texttt{NextCoder-7B}, showed improved accuracy when instructed to reason. In general, the low accuracies were expected, as the authors of the  LongBench~\cite{longBench} also reported similarly low results.
\par Figure~\ref{fig:codeLlamaTokenGen} illustrates a snapshot of the token generation process during a single inference and depicts the energy consumption for each generated token. It initially exhibits a sharp spike, followed by a sudden drop and a gradual increase with each newly generated token. This pattern corresponds to the inference phases where the prefill starts and decoding follows. Since prefill processes the entire prompt and initializes the key-value cache, its generated token is more energy-intensive than any token produced during decoding. However, we observed that generating the first token in the decoding phase (the second token overall) can also be energy-intensive, depending on the input length. This behavior likely stems from the transition between the prefill and decoding phases, where cache initialization, synchronization overheads, and hardware-level adjustments temporarily increase latency and energy consumption before the generation process stabilizes. 
\par Although the plot corresponds to \texttt{CodeLlama-7B} performing \texttt{0-shot CoT} generation, the same general pattern appears across all models and workloads. However, the magnitude of the initial spike and the subsequent rate of increase vary depending on the model, input size, and output length. Each of these aspects are discussed in the following subsections. \hl{Even though the accuracy of the model is important, the focus of our work is on the energy consumption of its inference.} Therefore, the focus in the next subsections is on energy efficiency.
\tablePrefillModels

\subsection{About input size}
\begin{figure}[t!]
    \centering
    \includegraphics[width=\linewidth]{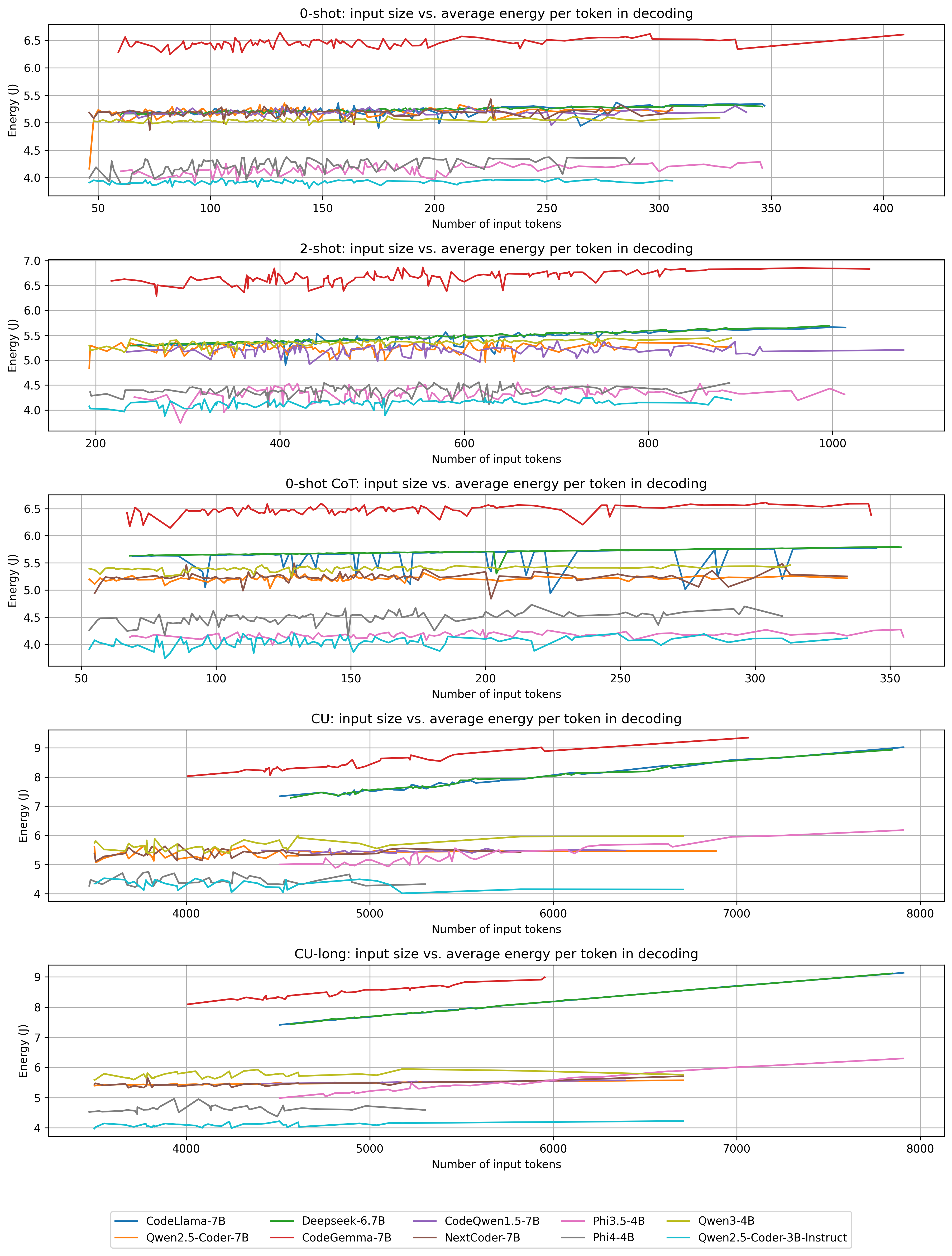}
    \caption{The plots illustrate the relationship between input size and energy consumption during the prefill phase (left) and per token in the decoding phase (right).}
    \label{fig:EnergyInputAll}
\end{figure}
\par \underline{\textbf{Impact on prefill.}} \hl{Table}~\ref{tab:resultsPrefill} \hl{presents the contribution of the prefill phase to the total energy consumption of the inference. For \texttt{0-shot} and \texttt{2-shot} the maximum output length was fixed at 300 tokens, with the \texttt{2-shot} input being approximately three times larger than the \texttt{0-shot} input. Across all models, increasing the input size led to an increase of 0.4\%–2.5\% in contribution of the prefill. To examine a more substantial increase, we compare \texttt{0-shot} and \texttt{CU-long}, where the maximum token limit is the same but the input length for \texttt{CU-long} ranges from 4,000 to 8,000 tokens. The increase in contribution now ranges between 4.7\% and 21.6\%. This increase is expected, since a larger input length leads to higher-dimensional internal matrices and to an increase in computational workload. }
\par \hl{Nevertheless, the prefill contribution remains much lower than decoding, except when the input size substantially exceeds the output length, as in the \texttt{CU} setting with 4,000–8,000 input tokens and a maximum output length of 10. Under this scenario, the contribution varies between 67.3\% and 84.4\%, whereas in all other cases it does not exceed 23\%. These results indicate that the decoding phase remains the primary contributor to overall energy consumption.}
\tableTokenizers
\tableDecodingModels
\par Lastly, the input size noticeably influences the energy-per-token metric, as shown in Table~\ref{tab:resultsTokenCosts}. When comparing two workloads with different input sizes but the same maximum output length, such as \texttt{0-shot} and \texttt{CU-long}, we observe that the energy per token varies for the same model. For instance, \texttt{Deepseek-6.7B} produced outputs of similar length for both workloads, but the average energy cost per token escalated from 5.22 to 8.59, showing a 64.5\% increase. Overall, the models exhibit varying sensitivity to increases in input size. One possible explanation lies in the tokenizers used by different models, which means that the same prompt can be broken down into varying numbers of tokens, as shown in Table~\ref{tab:tokenizers}. For instance, \texttt{CodeLlama-7B} tokenizes the prompts from the \texttt{0-shot} workload into an average of 163 tokens, whereas \texttt{Qwen2.5-Coder-7B} uses 137 tokens. The difference becomes more pronounced for the \texttt{CU} and \texttt{CU-long} workloads, where \texttt{CodeLlama-7B} uses 5,555 tokens compared to 4,287 for \texttt{Qwen2.5-Coder-7B}, representing a 30\% shorter input for the latter.\\

\begin{figure*}
    \centering
    \includegraphics[width=\linewidth]{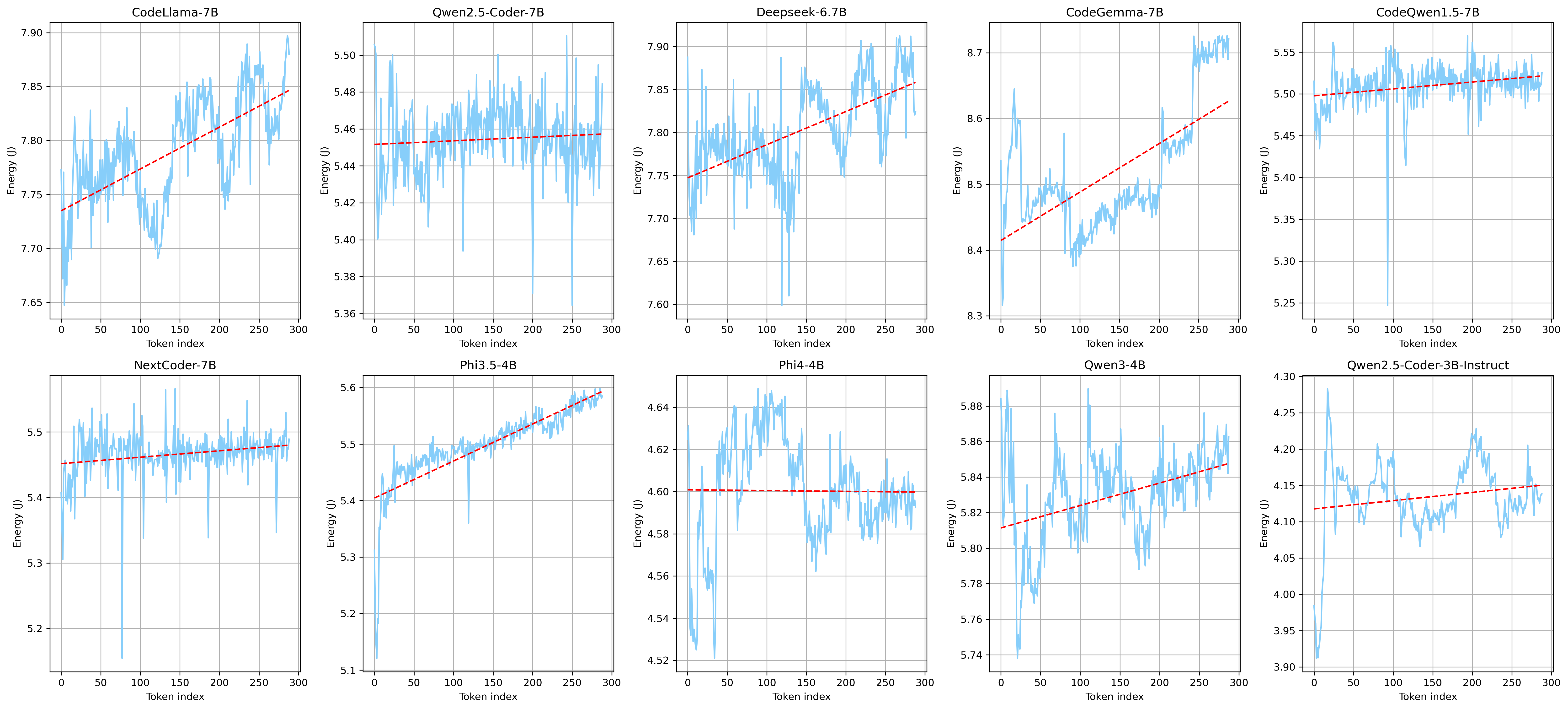}
    \caption{The plots illustrate token generation during inference, with the x-axis representing the index of each generated token and the y-axis representing the energy consumption for that token. The examples correspond to code understanding with long output.}
    \label{fig:TokenGenerationCULong}
\end{figure*}
\begin{figure*}
    \centering
    \includegraphics[width=\linewidth]{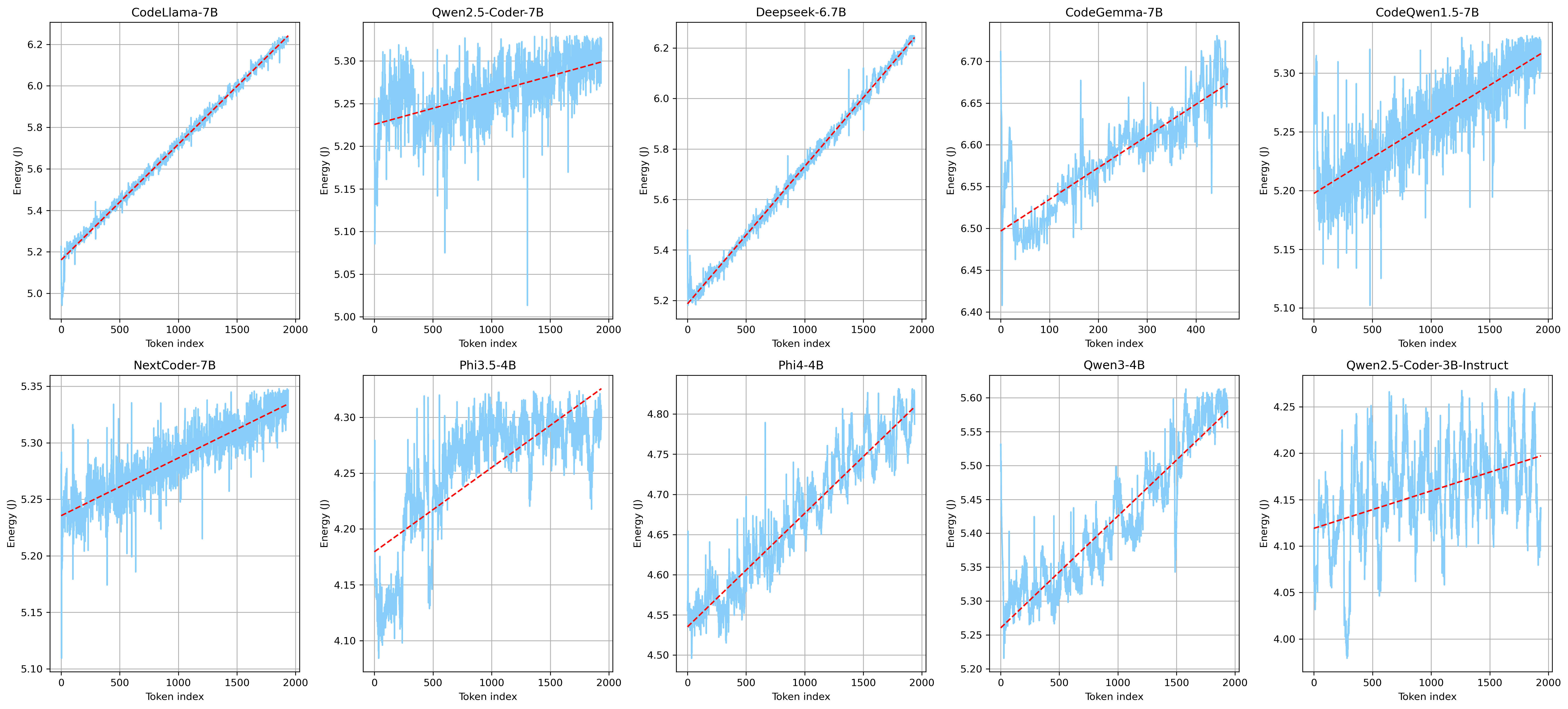}
    \caption{The plots illustrate token generation during inference, with the x-axis representing the index of each generated token and the y-axis representing the energy consumption for that token. The examples shown correspond to \texttt{0-shot} prompting with chain-of-thought.}
    \label{fig:TokenGenerationCoT}
\end{figure*}
\par \underline{\textbf{Impact on decoding.}} The plots on Figure~\ref{fig:EnergyInputAll} show the relationship between input size and the mean per-token energy cost in the decoding phase across all workloads and models. For the \texttt{0-shot} and \texttt{2-shot} workloads, none of the models exhibit a steep increase in token cost. \hl{However, as input size grows, as in \texttt{CU} and \texttt{CU-long}, four out of ten models show an evident rise in decoding token cost. \texttt{CodeGemma-7B}, \texttt{CodeLlama-7B}, and \texttt{Deepseek-6.7B} exhibit a noticeable increase in energy consumption for the \texttt{CU} and \texttt{CU-long} workloads, where the input size ranges between 4000 and 8000 tokens. \texttt{Phi3.5-4B} also exhibits an interesting pattern: it begins as one of the least energy-consuming models but, after exceeding 6500 input tokens, becomes the fourth most energy-intensive model in terms of average energy per token during decoding. This shows that the input size affects some models far more than others.}
\par Table~\ref{tab:resultsTokenCosts} presents the mean energy consumption per token during the decoding phase for all models and workloads, along with the corresponding standard deviations. For models such as \texttt{CodeLlama-7B}, \texttt{Deepseek-6.7B}, \texttt{CodeGemma-7B}, and \texttt{Phi3.5-4B}, the increase in per-token decoding cost is more pronounced than in other models when moving from the \texttt{0-shot} to the \texttt{CU-long} workload, where the input size increases by at least an order of magnitude while the maximum output length remains the same. This indicates that input size can influence the energy cost of tokens during the decoding.
\par From our experiments, we observed that once the prefill phase is completed, it establishes the initial energy cost for tokens in the decoding phase, which is particularly evident in the models listed above. Figure~\ref{fig:TokenGenerationCULong} illustrates token generation across all models for \texttt{CU-long}, while Figure~\ref{fig:TokenGenerationCoT} presents the same for \texttt{0-shot CoT}. The plots include regression lines, in red, to show the progression of energy consumption for each newly generated token. From both figures, we see that the initial decoding cost is lower for \texttt{0-shot CoT} compared to \texttt{CU-long}, which can be attributed to the smaller input size. For example, based on the linear fit for \texttt{CodeLlama-7B}, the decoding begins at approximately 5.2J for \texttt{0-shot CoT}, compared to around 7.74J for \texttt{CU-long}, showing a 48.8\% increase. This trend is consistent across all models, but the degree of variation differs among them. A possible explanation is that the prefill phase initializes the key-value cache, whose size depends on the input length. Hence, a larger initial key-value cache results in a higher initial decoding cost, as a bigger matrix entails more computations.
\par However, not all models exhibit a significant increase in energy per token during decoding as the input length grows. For instance, \texttt{Phi4-4B} starts its decoding at 4.54J for \texttt{0-shot CoT} and at 4.60J for \texttt{CU-long}, corresponding to only a 1.3\% increase. Such differences may arise from architectural variations between models, including the use of sparse or efficient attention mechanisms, as well as more optimized storage of key-value caches.
\begin{summary}
{\footnotesize 
\textbf{Summary.} 
\hl{The results show that in four out of five scenarios, the decoding phase dominates energy consumption, except in one case where the input length significantly exceeds the output length. Furthermore, the input length affects both inference phases. As the input length increases, prefill energy rises, and tokens in the decoding phase also become more energy-intensive. Additionally, the initial energy per token during decoding increases. On the one hand, this increase can be significant, reaching 51.8\% for \texttt{CodeLlama-7B}. On the other hand, the magnitude of the increase depends on the model and can vary between 1.3\% and 51.8\%.}
\\ \textbf{Implications.} \hl{From the perspective of LLM implementation, inference optimization strategies should account for the fact that the prefill phase can influence energy consumption of the decoding phase. In particular, efficient storage and retrieval of the key–value cache can help mitigate this effect. From the perspective of LLM users, increasing input length raises the overall cost of inference. Therefore, the prompt should be increased mindfully and only when necessary.}
}
\end{summary}
\subsection{Output length \& Decoding}
\par Table~\ref{tab:resultsTotal} presents the average total energy consumption per inference alongside the average output length for each model and workload. Models producing the longest outputs on average, such as \texttt{CodeLlama-7B}, \texttt{Deepseek-6.7B}, \texttt{CodeQwen1.5-7B}, and \texttt{Qwen3-4B} exhibit the highest total energy consumption.
\par A more interesting phenomenon is that each newly generated token tends to be more energy-intensive than the previous one. Figure~\ref{fig:TokenGenerationCULong} and Figure \ref{fig:TokenGenerationCoT} show the energy consumption for each token during generation for \texttt{CU-long} and \texttt{0-shot CoT}. The values are averaged across the entire benchmark, so the plots provide a snapshot of the token-level energy behavior for each model and workload. In Figure~\ref{fig:TokenGenerationCULong}, the maximum output length was set to 300 tokens, and it is already possible to observe an increase in energy consumption from the first to the last generated token for some models. For instance, in \texttt{CodeLlama-7B}, the first token consumes approximately 7.74J, while the last token reaches 7.89J. Similarly, for \texttt{Phi3.5-4B}, the first token consumes 5.4J, increasing to 5.6J for the last token. The increasing trend also applies to \texttt{Deepseek-6.7B}, \texttt{CodeGemma-7B} and \texttt{Qwen3-4B}. 
\par On the other hand, Figure~\ref{fig:TokenGenerationCoT} shows token generation for \texttt{0-shot CoT}, with the maximum output length set to 2000 tokens. Generation stops either when the model produces the end-of-sentence token or reaches the set limit. Not every model reached this maximum length. Thus, models such as \texttt{Qwen2.5-Coder-7B}, \texttt{CodeGemma-7B}, and \texttt{CodeQwen1.5-7B} reached the end-of-sentence token before producing 2000 tokens. Figure~\ref{fig:TokenGenerationCoT} highlights a more pronounced trend in the increase in energy per generated token across all models, although the magnitude of the increase varies by model. For instance, the rise is sharper for \texttt{CodeLlama-7B} and \texttt{Deepseek-6.7B}, with an increase of approximately 20\% in the energy cost of the last token compared to the first in the decoding phase. Other notable increases are seen in \texttt{NextCoder-7B} with 11\%, \texttt{Phi4-4B} with 5\%, and \texttt{Qwen3-4B} with 7\%. The other models also show an increase in energy consumption with each newly generated token. \hl{However, the rate of this increase is not as significant as for the models mentioned above.}
\par The phenomenon of each newly generated token being more expensive than the previous one can arise from several factors. In transformers, for example, each token incorporates information from all preceding tokens by computing weighted relationships through self-attention. As the sequence grows, the attention matrix becomes larger, leading to increased computations per token, which is particularly noticeable in long outputs. Additionally, generating a new token requires updating the key-value cache and sometimes reorganizing memory, which adds further overhead. Hardware and runtime factors, such as memory allocation, cache misses, or kernel launches, can also contribute to higher energy consumption per token as the sequence length increases. Hence, as we observed different models exhibit varying patterns of increase, and the exact cause of this phenomenon depends on each model’s specific implementation and behavior.
\begin{summary}
{\footnotesize 
\textbf{Summary.} 
The results indicate that token generation becomes increasingly expensive with each successive token. The difference can be substantial, with the last token in a 2000-token output sequence costing up to 20\% more than the first. This phenomenon can arise from factors such as the growing key-value cache and memory management overhead. However, the exact cause depends on the model’s implementation, as the pattern and magnitude of the increase vary between models. Also, some models, such as \texttt{Deepseek-6.7B}, \texttt{CodeLlama-7B}, and \texttt{Qwen3-4B}, tend to generate outputs close to the maximum token limit, even when doing so is unnecessary for successful task completion.
\\ \textbf{Implications.} Since each successive token is more expensive to generate than the previous one, it is important to limit output length whenever possible to avoid unnecessary energy consumption. Implementing early stopping criteria can substantially reduce energy usage without affecting task performance, as some models continue generating whitespaces or formatting characters even after successfully completing a task.
}
\end{summary}

\section{Efficient code generation via \textit{babbling suppression}}

\par  In our experiments, we observed that some models tend to generate outputs close to the maximum allowed length. For example, \texttt{Deepseek-6.7B} produced an average of 300 tokens when the limit was set to 300, and 1989 tokens on average when the limit was 2000. Similarly, \texttt{CodeLlama-7B} and \texttt{Qwen3-4B} also tended to generate outputs near the maximum limit. After examining the outputs, it was observed that after generating the target function, those models continued producing additional text such as white spaces, test cases, usage examples, or alternative implementations. We call these models \textit{babblers} since, according to the Merriam–Webster dictionary, \textit{babbling} is \textit{“the production of meaningless strings of speech"}~\cite{babble}.
\par Based on our findings, the decoding phase remains the dominant contributor to the energy cost of LLM inference. The generation of each new token incrementally increases the overall energy consumption. Thus, the most straightforward approach to improving energy efficiency is to minimize the number of generated tokens. In the context of code generation, the most energy-efficient strategy is to produce only the required function, without any extra text, unless it is explicitly requested. However, babblers incur unnecessary energy overhead, as they produce content beyond what is required for a correct solution. During post-processing, only the generated function is extracted, while all additional text is discarded, resulting in avoidable energy waste.
\begin{algorithm}
\caption{Code Generation with Babbling Suppression}
\label{alg:incremental_generation}
\begin{algorithmic}[1]
\Require LLM $M$, Prompt $P$, Test Cases $T$
\State $generated\_code \gets ""$
\While{generation not terminated}
    \State $t \gets \Call{GenerateToken}{M, P}$
    \State $generated\_code \gets generated\_code++t$
    \If{$t$ is end-of-line token}
        \If{\Call{Compiles}{generated\_code}}
            \If{\Call{PassesTests}{generated\_code, T}}
                \State \Return $generated\_code$ 
            \EndIf
        \EndIf
    \EndIf
\EndWhile
\State \Return $generated\_code$ 
\end{algorithmic}
\end{algorithm}
\tableEarlyStopping
\par To prevent models from babbling, we leverage three observations: (i) in our experiments, most of the extraneous tokens are produced \textit{after} the actual solutions, (ii) for code, it is possible to objectively verify generated solutions by testing, and (iii) checking whether solutions are correct is much cheaper than generating them. Thus, stopping early can be a cost-effective way of improving energy efficiency in code generation. Algorithm~\ref{alg:incremental_generation} shows how we have implemented this approach, that we call \textit{babbling suppression}. After each newly generated token, we check whether it corresponds to an end-of-line token. This per-line approach reduces the number of syntactic and test checks, avoiding the need to perform them after every token. If the line is completed, then the partially generated output is checked for syntactic correctness. If the code executes without errors, then the corresponding benchmark test cases are run. Generation is halted as soon as all test cases pass. This algorithm operates externally and does not influence the model’s internal behavior.

\par Table~\ref{tab:earlystopping} presents the results of applying \textit{babbling suppression} algorithm to babbling models. The evaluation considers two configurations: one in which the maximum token limit is set to 300 tokens, matching the original \texttt{0-shot} setting, and another in which the limit is increased to 1000 tokens. The latter configuration allows us to examine whether babblers attempt to exhaust the token budget. The results indicate that the \textit{babbling suppression} consistently halts generation earlier than the baseline across all models, without a significant impact on accuracy. Owing to the non-deterministic nature of LLMs, minor variations in accuracy, $\pm 2\%$, are expected. 
\par With a maximum of 300 tokens, \textit{babbling suppression} achieves at least a 44\% reduction in output length. The largest savings are observed for \texttt{Deepseek-6.7B}, with an 89\% reduction in output tokens, resulting in a 69\% reduction in energy consumption per inference. This model also tends to exhaust the token budget when the limit is increased to 1000 tokens. Applying \textit{babbling suppression} in this setting yields a 93\% reduction in output tokens without affecting accuracy, resulting in a 89\% reduction in energy consumption.
\par However, our approach introduces some overhead due to test checks on intermediate outputs, which is reflected in the energy-per-token metric. A notable example is \texttt{CodeLlama-7B} with a 300-token limit, where inference energy increases by 6\% and energy per token by 53\%. Nevertheless, this overhead negatively affected the mean energy consumption of inference only in this one case, and it can be mitigated by further reducing the number of checks, as the current approach performs them on a per-line basis. 
It could be argued that reducing the token limit would have similar results. However, the consequences are not obvious. For instance, increasing the token limit for \texttt{Qwen3-4B} from 300 to 1000 raised accuracy from 40\% to 68\%. Moreover, the problem of unnecessarily generated test cases, usage examples, or whitespace potentially remains. These "leftovers" are typically removed via post-processing, meaning that the energy spent on generating them is wasted.

\begin{summary}
{\footnotesize 
\textbf{Summary.} 
We implemented a \textit{babbling suppression} to control the \textit{babbling} behavior of \texttt{CodeLlama-7B}, \texttt{Deepseek-6.7B}, and \texttt{Qwen3-4B} by introducing test checks on intermediate outputs. Reduction in the number of generated tokens ranged from 44\% (300-token limit) to 93\% (1000-token limit). In terms of energy per inference, the reduction ranged from 44\% (300-token limit) to 89\% (1000-token limit). Some overhead is introduced by intermediate test checks, but it negatively affected inference energy in only one of six cases, with a 6\% increase.
\\ \textbf{Implications.} \textit{Babbling suppression} effectively prevents excessive token generation, reducing energy consumption during code generation by eliminating unnecessary tokens that would otherwise be discarded during post-processing. 
}
\end{summary}

\section{Threats to validity}
\par \textbf{Construct validity.} In this study, we aim to measure energy consumption at the phase level of LLM inference. \hl{A potential threat to construct validity arises from the definition of phase boundaries, specifically whether the generation of the first token is included in the prefill phase or considered part of the decoding phase. Consistent with multiple studies}~\cite{splitwise, spad, distserve}, \hl{we include the first token in the prefill phase.} Another potential threat to construct validity is that, due to the high speed of inference, measurements may not perfectly align with a single phase, potentially skewing per-phase energy calculations. To mitigate this, we employed the methodology originally proposed by Babakol et al.~\cite{methodology_babakol}, periodically recording energy measurements with timestamps and aligning them with the timestamps of token generation. More specifically, the energy measurement for token \textit{n} is used if its timestamp falls between the timestamp for token \textit{n-1} and timestamp for token \textit{n}, with all token timestamps being recorded at the end of their generation.

\par \textbf{Internal validity.} As the experiments were conducted on UT JupyterLab, which is a shared environment accessible to all university employees, exclusive GPU access could not be ensured. This introduces a possible threat that other users’ activity on the same GPU could affect the energy measurements. To prevent that, we monitor GPU utilization prior to each inference run. We also verify that only one process is active on the GPU, allowing the inference to proceed only when exclusive GPU access is confirmed. There is also a potential threat to the evaluation of the accuracy, as some models generate additional text besides the target function, either preceding or following it. Nevertheless, we parsed each response and extracted all code snippets generated by the models, verifying that they were complete and adhered to Python syntax and structure. Despite the efforts, there is a chance that this process still negatively influenced the models' accuracy. 
\par \textbf{External validity.} Our findings are based on a single model architecture - decoder-only transformers. Therefore, they may not necessarily generalize to other architectures, such as encoder-only transformers, encoder-decoder transformers or mixture-of-experts models. Nevertheless, decoder-based transformers represent the most widely adopted class of LLMs, as evidenced by popular families such as GPT. Furthermore, our study includes ten models of varying sizes and from different model families, providing a diverse and representative sample within this architecture type. 
\par \textbf{Conclusion validity.} 
A potential threat to conclusion validity lies in the risk of drawing incorrect results due to limited sample sizes, data variability, or inappropriate statistical analysis. In this study, we used two datasets: HumanEval, consisting of 164 data entries applied across three workloads, and LongBench, containing 50 entries used for two workloads. To enhance the reliability of our results, we also removed outliers from the data to ensure that the computed mean values accurately represent the overall trends. Lastly, we report standard deviation, to add the credibility to the mean values.  
\section{Conclusion}
\par \hl{The primary goal of this study is to evaluate the energy consumption of the prefill and decoding phases of LLM inference and assess their contributions to the total cost. Unlike existing studies, which typically measure energy consumption over the entire output, our measurements are conducted at the per-token level.
Our results show that decoding generally dominates energy consumption in LLM inference, and longer inputs increase energy costs in both prefill and decoding phases. Token generation becomes progressively more expensive, with the last tokens being up to 20\% more expensive than the first one. Three out of ten models demonstrate \textit{babbling} behavior, generating unnecessary tokens. This is mitigated by using a \textit{babbling suppression} algorithm that halts generation once all tests pass. This approach can reduce inference energy consumption by up to 89\% by preventing the generation of unnecessary tokens that are typically removed during post-processing.}
\par \textbf{Implications.} 
\hl{These findings highlight the importance of optimizing both input and output handling in LLM inference. From an implementation perspective, prefill-phase design, such as  efficient KV cache storage and retrieval, can influence decoding energy. From a user perspective, prompt length and output length should be carefully controlled, as both directly impact inference cost. Limiting unnecessary token generation through \textit{babbling suppression} can substantially reduce energy consumption without degrading task performance. This prevents wasted computation on outputs that would otherwise be discarded during post-processing.}

\bibliographystyle{ACM-Reference-Format}
\bibliography{sample-base}


\end{document}

%% file: tables.tex
\newcommand{\tableResultsModels}{%
\begin{table*}[t!]
\centering
\tiny
\begin{tabular}{|lccccccccccccccc|}
\hline
\multicolumn{1}{|c|}{\multirow{2}{*}{\textbf{Model}}} & \multicolumn{5}{c|}{\textbf{Accuracy (\%)}}                                                                                                                                           & \multicolumn{5}{c|}{\textbf{Output}}                                                                                                                                                              & \multicolumn{5}{c|}{\textbf{Total Energy (J)}}                                                                                                                                \\ \cline{2-16} 
\multicolumn{1}{|c|}{}                                & \multicolumn{1}{c|}{\textit{0-shot}} & \multicolumn{1}{c|}{\textit{2-shot}} & \multicolumn{1}{c|}{\textit{0-shot CoT}} & \multicolumn{1}{c|}{\textit{CU}} & \multicolumn{1}{l|}{\textit{CU-long}} & \multicolumn{1}{c|}{\textit{0-shot}} & \multicolumn{1}{c|}{\textit{2-shot}} & \multicolumn{1}{c|}{\textit{0-shot CoT}} & \multicolumn{1}{c|}{\textit{CU}} & \multicolumn{1}{c|}{\textit{CU-long}} & \multicolumn{1}{c|}{\textit{0-shot}} & \multicolumn{1}{c|}{\textit{2-shot}} & \multicolumn{1}{c|}{\textit{0-shot CoT}} & \multicolumn{1}{c|}{\textit{CU}}  & \textit{CU-long} \\ \hline
\multicolumn{16}{|c|}{\textbf{6-7B Group}}                                                                                                                                                                                                                                                                                                                                                                                                                                                                                                                                                                                        \\ \hline
\multicolumn{1}{|l|}{CodeLlama-7B}                    & \multicolumn{1}{c|}{34}              & \multicolumn{1}{c|}{35}              & \multicolumn{1}{c|}{26}                  & \multicolumn{1}{c|}{6}            & \multicolumn{1}{c|}{6}   & \multicolumn{1}{c|}{$207(\pm120)$}   & \multicolumn{1}{c|}{$248(\pm102)$}   & \multicolumn{1}{c|}{$1765(\pm622)$}      & \multicolumn{1}{c|}{$10$}        & \multicolumn{1}{c|}{$300(\pm0)$}            & \multicolumn{1}{c|}{$1082 (\pm632)$} & \multicolumn{1}{c|}{$1369(\pm559)$}  & \multicolumn{1}{c|}{$10069(\pm3569)$}    & \multicolumn{1}{c|}{$258(\pm36)$} & $2582(\pm143)$   \\ \hline
\multicolumn{1}{|l|}{Qwen2.5-Coder-7B}                & \multicolumn{1}{c|}{69}              & \multicolumn{1}{c|}{69}              & \multicolumn{1}{c|}{74}                  & \multicolumn{1}{c|}{34}           & \multicolumn{1}{c|}{36}   & \multicolumn{1}{c|}{$107(\pm73)$}    & \multicolumn{1}{c|}{$97(\pm83)$}     & \multicolumn{1}{c|}{$107(\pm78)$}        & \multicolumn{1}{c|}{$10$}        & \multicolumn{1}{c|}{$279(\pm67)$}     & \multicolumn{1}{c|}{$575(\pm391)$}   & \multicolumn{1}{c|}{$541(\pm456)$}   & \multicolumn{1}{c|}{$576(\pm421)$}       & \multicolumn{1}{c|}{$188(\pm24)$} & $1661(\pm366)$   \\ \hline
\multicolumn{1}{|l|}{Deepseek-Coder-6.7B}             & \multicolumn{1}{c|}{61}              & \multicolumn{1}{c|}{66}              & \multicolumn{1}{c|}{56}                  & \multicolumn{1}{c|}{14}           & \multicolumn{1}{c|}{16}   & \multicolumn{1}{c|}{$300(\pm0)$}           & \multicolumn{1}{c|}{$300(\pm0)$}           & \multicolumn{1}{c|}{$1989(\pm132)$}      & \multicolumn{1}{c|}{$10$}        & \multicolumn{1}{c|}{$300(\pm0)$}            & \multicolumn{1}{c|}{$1567(\pm13)$}   & \multicolumn{1}{c|}{$1664(\pm37)$}   & \multicolumn{1}{c|}{$11338(\pm772)$}     & \multicolumn{1}{c|}{$265(\pm31)$} & $2579(\pm133)$   \\ \hline
\multicolumn{1}{|l|}{CodeGemma-7B}                    & \multicolumn{1}{c|}{54}              & \multicolumn{1}{c|}{49}              & \multicolumn{1}{c|}{53}                  & \multicolumn{1}{c|}{20}           & \multicolumn{1}{c|}{24}   & \multicolumn{1}{c|}{$86(\pm60)$}     & \multicolumn{1}{c|}{$117(\pm84)$}    & \multicolumn{1}{c|}{$101(\pm68)$}        & \multicolumn{1}{c|}{$10$}        & \multicolumn{1}{c|}{$80(\pm78)$}      & \multicolumn{1}{c|}{$553(\pm393)$}   & \multicolumn{1}{c|}{$797(\pm569)$}   & \multicolumn{1}{c|}{$654(\pm445)$}       & \multicolumn{1}{c|}{$270(\pm30)$} & $870(\pm672)$    \\ \hline
\multicolumn{1}{|l|}{CodeQwen1.5-7B}                  & \multicolumn{1}{c|}{45}              & \multicolumn{1}{c|}{48}              & \multicolumn{1}{c|}{42}                  & \multicolumn{1}{c|}{4}            & \multicolumn{1}{c|}{2}   & \multicolumn{1}{c|}{$155(\pm118)$}   & \multicolumn{1}{c|}{$112(\pm102)$}   & \multicolumn{1}{c|}{$150(\pm244)$}       & \multicolumn{1}{c|}{$10$}        & \multicolumn{1}{c|}{$300(\pm0)$}            & \multicolumn{1}{c|}{$805(\pm608)$}   & \multicolumn{1}{c|}{$628(\pm561)$}   & \multicolumn{1}{c|}{$823(\pm1377)$}      & \multicolumn{1}{c|}{$230(\pm18)$} & $1834(\pm27)$    \\ \hline
\multicolumn{1}{|l|}{NextCoder-7B}                    & \multicolumn{1}{c|}{40}              & \multicolumn{1}{c|}{11}              & \multicolumn{1}{c|}{63}                  & \multicolumn{1}{c|}{34}           & \multicolumn{1}{c|}{32}   & \multicolumn{1}{c|}{$160(\pm125)$}   & \multicolumn{1}{c|}{$274(\pm76)$}    & \multicolumn{1}{c|}{$321(\pm537)$}       & \multicolumn{1}{c|}{$10$}        & \multicolumn{1}{c|}{$104(\pm121)$}    & \multicolumn{1}{c|}{$831(\pm648)$}   & \multicolumn{1}{c|}{$1448(\pm401)$}  & \multicolumn{1}{c|}{$1780(\pm3012)$}     & \multicolumn{1}{c|}{$187(\pm23)$} & $708(\pm658)$    \\ \hline
\multicolumn{16}{|c|}{\textbf{3-4B Group}}                                                                                                                                                                                                                                                                                                                                                                                                                                                                                                                                                                                        \\ \hline
\multicolumn{1}{|l|}{Phi3.5-4B}                       & \multicolumn{1}{c|}{70}              & \multicolumn{1}{c|}{66}              & \multicolumn{1}{c|}{68}                  & \multicolumn{1}{c|}{14}           & \multicolumn{1}{c|}{18}   & \multicolumn{1}{c|}{$161(\pm78)$}    & \multicolumn{1}{c|}{$125(\pm95)$}    & \multicolumn{1}{c|}{$262(\pm94)$}        & \multicolumn{1}{c|}{$10$}        & \multicolumn{1}{c|}{$250(\pm91)$}     & \multicolumn{1}{c|}{$670(\pm327)$}   & \multicolumn{1}{c|}{$531(\pm400)$}   & \multicolumn{1}{c|}{$1144(\pm408)$}      & \multicolumn{1}{c|}{$170(\pm22)$} & $1450(\pm501)$   \\ \hline
\multicolumn{1}{|l|}{Phi4-4B}                         & \multicolumn{1}{c|}{75}              & \multicolumn{1}{c|}{54}              & \multicolumn{1}{c|}{75}                  & \multicolumn{1}{c|}{16}           & \multicolumn{1}{c|}{16}   & \multicolumn{1}{c|}{$132(\pm79)$}    & \multicolumn{1}{c|}{$99(\pm76)$}     & \multicolumn{1}{c|}{$214(\pm314)$}       & \multicolumn{1}{c|}{$10$}        & \multicolumn{1}{c|}{$149(\pm107)$}    & \multicolumn{1}{c|}{$560(\pm341)$}   & \multicolumn{1}{c|}{$431(\pm330)$}   & \multicolumn{1}{c|}{$976(\pm1462)$}      & \multicolumn{1}{c|}{$111(\pm10)$} & $768(\pm503)$    \\ \hline
\multicolumn{1}{|l|}{Qwen3-4B}                        & \multicolumn{1}{c|}{40}              & \multicolumn{1}{c|}{19}              & \multicolumn{1}{c|}{76}                  & \multicolumn{1}{c|}{22}           & \multicolumn{1}{c|}{22}   & \multicolumn{1}{c|}{$299(\pm1)$}     & \multicolumn{1}{c|}{$299(\pm3)$}     & \multicolumn{1}{c|}{$1819(\pm420)$}      & \multicolumn{1}{c|}{$10$}        & \multicolumn{1}{c|}{$299(\pm1)$}      & \multicolumn{1}{c|}{$1508(\pm11)$}   & \multicolumn{1}{c|}{$1606(\pm59)$}   & \multicolumn{1}{c|}{$9853(\pm2303)$}     & \multicolumn{1}{c|}{$144(\pm19)$} & $1812(\pm39)$    \\ \hline
\multicolumn{1}{|l|}{Qwen2.5-Coder-3B}                & \multicolumn{1}{c|}{63}              & \multicolumn{1}{c|}{65}              & \multicolumn{1}{c|}{69}                  & \multicolumn{1}{c|}{22}           & \multicolumn{1}{c|}{26}   & \multicolumn{1}{c|}{$260(\pm58)$}    & \multicolumn{1}{c|}{$241(\pm91)$}    & \multicolumn{1}{c|}{$82(\pm57)$}         & \multicolumn{1}{c|}{$10$}        & \multicolumn{1}{c|}{$275(\pm62)$}     & \multicolumn{1}{c|}{$1021(\pm229$)}  & \multicolumn{1}{c|}{$1000(\pm376)$}  & \multicolumn{1}{c|}{$334(\pm230)$}       & \multicolumn{1}{c|}{$100(\pm10)$} & $1218(\pm261)$   \\ \hline
\end{tabular}
\caption{Results show the mean and standard deviation of pass@1 accuracy (\%), output (in number of tokens), and total energy consumption (in joules) per inference for each LLM. These statistics are computed across the benchmark. }
\label{tab:resultsTotal}
\end{table*}%
}

\newcommand{\tableDecodingModels}{%
\begin{table*}[ht!]
\centering
\scriptsize
\begin{tabular}{|lcccccccccc|}
\hline
\multicolumn{1}{|c|}{\multirow{2}{*}{\textbf{Model}}} & \multicolumn{5}{c|}{\textbf{Energy per Token (J)}}                                                                                                                                                      & \multicolumn{5}{c|}{\textbf{Energy per Token in Decoding (J)}}                                                                                                                   \\ \cline{2-11} 
\multicolumn{1}{|c|}{}                                & \multicolumn{1}{c|}{\textit{0-shot}} & \multicolumn{1}{c|}{\textit{2-shot}} & \multicolumn{1}{c|}{\textit{0-shot CoT}} & \multicolumn{1}{c|}{\textit{CU}}      & \multicolumn{1}{c|}{\textit{CU-long}}  & \multicolumn{1}{c|}{\textit{0-shot}} & \multicolumn{1}{c|}{\textit{2-shot}} & \multicolumn{1}{c|}{\textit{0-shot CoT}} & \multicolumn{1}{c|}{\textit{CU}}     & \textit{CU-long} \\ \hline
\multicolumn{11}{|c|}{\textbf{6-7B Group}}                                                                                                                                                                                                                                                                                                                                                                                                         \\ \hline
\multicolumn{1}{|l|}{CodeLlama-7B}                    & \multicolumn{1}{c|}{$5.24(\pm0.09)$} & \multicolumn{1}{c|}{$5.58(\pm0.21)$} & \multicolumn{1}{c|}{$5.65(\pm0.17)$}     & \multicolumn{1}{c|}{$25.77(\pm3.56)$} & \multicolumn{1}{c|}{$8.61(\pm0.48)$}   & \multicolumn{1}{c|}{$5.21(\pm0.09)$} & \multicolumn{1}{c|}{$5.44(\pm0.12)$} & \multicolumn{1}{c|}{$5.64(\pm0.19)$}     & \multicolumn{1}{c|}{$7.82(\pm0.38)$} & $7.97(\pm0.38)$  \\ \hline
\multicolumn{1}{|l|}{Qwen2.5-Coder-7B}                & \multicolumn{1}{c|}{$5.34(\pm0.09)$} & \multicolumn{1}{c|}{$5.73(\pm0.38)$} & \multicolumn{1}{c|}{$5.38(\pm0.09)$}     & \multicolumn{1}{c|}{$19.39(\pm3.39)$} & \multicolumn{1}{c|}{$6.81(\pm4.72)$}   & \multicolumn{1}{c|}{$5.31(\pm0.08)$} & \multicolumn{1}{c|}{$5.43(\pm0.14)$} & \multicolumn{1}{c|}{$5.33(\pm0.08)$}     & \multicolumn{1}{c|}{$5.37\pm0.16)$}  & $5.46(\pm0.05)$  \\ \hline
\multicolumn{1}{|l|}{Deepseek-Coder-6.7B}             & \multicolumn{1}{c|}{$5.22(\pm0.04)$} & \multicolumn{1}{c|}{$5.54(\pm0.12)$} & \multicolumn{1}{c|}{$5.72(\pm0.05)$}     & \multicolumn{1}{c|}{$26.48(\pm3.78)$} & \multicolumn{1}{c|}{$8.59(\pm0.44)$}   & \multicolumn{1}{c|}{$5.22(\pm0.04)$} & \multicolumn{1}{c|}{$5.48(\pm0.09)$} & \multicolumn{1}{c|}{$5.72(\pm0.05)$)}    & \multicolumn{1}{c|}{$7.82(\pm0.36)$} & $7.97(\pm0.35)$  \\ \hline
\multicolumn{1}{|l|}{CodeGemma-7B}                    & \multicolumn{1}{c|}{$6.45(\pm0.11)$} & \multicolumn{1}{c|}{$6.87(\pm0.24)$} & \multicolumn{1}{c|}{$6.46(\pm0.08)$}     & \multicolumn{1}{c|}{$27.59(\pm3.01)$} & \multicolumn{1}{c|}{$14.63(\pm6.14)$}  & \multicolumn{1}{c|}{$6.45(\pm0.08)$} & \multicolumn{1}{c|}{$6.65(\pm0.13)$} & \multicolumn{1}{c|}{$6.46(\pm0.07)$}     & \multicolumn{1}{c|}{$8.44(\pm0.28)$} & $8.51(\pm0.27)$  \\ \hline
\multicolumn{1}{|l|}{CodeQwen1.5-7B}                  & \multicolumn{1}{c|}{$5.25(\pm0.18)$} & \multicolumn{1}{c|}{$5.93(\pm0.85)$} & \multicolumn{1}{c|}{$5.47(\pm0.31)$}     & \multicolumn{1}{c|}{$22.99(\pm1.76)$} & \multicolumn{1}{c|}{$6.12(\pm0.09)$}   & \multicolumn{1}{c|}{$5.18(\pm0.06)$} & \multicolumn{1}{c|}{$5.45(\pm0.13)$} & \multicolumn{1}{c|}{$5.33(\pm0.14)$}     & \multicolumn{1}{c|}{$5.45(\pm0.04)$} & $5.52(\pm0.03)$  \\ \hline
\multicolumn{1}{|l|}{NextCoder-7B}                    & \multicolumn{1}{c|}{$5.25(\pm0.21)$} & \multicolumn{1}{c|}{$5.33(\pm0.19)$} & \multicolumn{1}{c|}{$5.46(\pm0.27)$}     & \multicolumn{1}{c|}{$21.09(\pm6.89)$} & \multicolumn{1}{c|}{$13.21(\pm10.62)$} & \multicolumn{1}{c|}{$5.18(\pm0.09)$} & \multicolumn{1}{c|}{$5.25(\pm0.05)$} & \multicolumn{1}{c|}{$5.34(\pm0.11)$}     & \multicolumn{1}{c|}{$5.42(\pm0.15)$} & $5.46(\pm0.11)$  \\ \hline
\multicolumn{11}{|c|}{\textbf{3-4B Group}}                                                                                                                                                                                                                                                                                                                                                                                                         \\ \hline
\multicolumn{1}{|l|}{Phi3.5-4B}                       & \multicolumn{1}{c|}{$4.15(\pm0.09)$} & \multicolumn{1}{c|}{$4.28(\pm0.17)$} & \multicolumn{1}{c|}{$4.36(\pm0.14)$}     & \multicolumn{1}{c|}{$17.23(\pm2.16)$} & \multicolumn{1}{c|}{$6.38(\pm1.72)$}   & \multicolumn{1}{c|}{$4.14(\pm0.09)$} & \multicolumn{1}{c|}{$4.15(\pm0.03)$} & \multicolumn{1}{c|}{$4.35(\pm0.14)$}     & \multicolumn{1}{c|}{$5.31(\pm0.32)$} & $5.44(\pm0.29)$  \\ \hline
\multicolumn{1}{|l|}{Phi4-4B}                         & \multicolumn{1}{c|}{$4.27(\pm0.13)$} & \multicolumn{1}{c|}{$4.42(\pm0.21)$} & \multicolumn{1}{c|}{$4.51(\pm0.14)$}     & \multicolumn{1}{c|}{$12.25(\pm1.38)$} & \multicolumn{1}{c|}{$6.39(\pm2.88)$}   & \multicolumn{1}{c|}{$4.22(\pm0.13)$} & \multicolumn{1}{c|}{$4.31(\pm0.11)$} & \multicolumn{1}{c|}{$4.49(\pm0.14)$}     & \multicolumn{1}{c|}{$4.45(\pm0.15)$} & $4.63(\pm0.11)$  \\ \hline
\multicolumn{1}{|l|}{Qwen3-4B}                        & \multicolumn{1}{c|}{$5.04(\pm0.03)$} & \multicolumn{1}{c|}{$5.37(\pm0.19)$} & \multicolumn{1}{c|}{$5.41(\pm0.05)$}     & \multicolumn{1}{c|}{$15.82(\pm2.12)$} & \multicolumn{1}{c|}{$6.06(\pm0.13)$}   & \multicolumn{1}{c|}{$5.04(\pm0.03)$} & \multicolumn{1}{c|}{$5.35(\pm0.18)$} & \multicolumn{1}{c|}{$5.41(\pm0.05)$}     & \multicolumn{1}{c|}{$5.65(\pm0.19)$} & $5.74(\pm0.09)$  \\ \hline
\multicolumn{1}{|l|}{Qwen2.5-Coder-3B}                & \multicolumn{1}{c|}{$3.93(\pm0.04)$} & \multicolumn{1}{c|}{$4.15(\pm0.07)$} & \multicolumn{1}{c|}{$4.06(\pm0.09)$}     & \multicolumn{1}{c|}{$11.01(\pm1.13)$} & \multicolumn{1}{c|}{$4.53(\pm0.62)$}   & \multicolumn{1}{c|}{$3.93(\pm0.15)$} & \multicolumn{1}{c|}{$4.02(\pm0.28)$} & \multicolumn{1}{c|}{$4.05(\pm0.09)$}     & \multicolumn{1}{c|}{$4.49(\pm0.19)$} & $4.22(\pm0.06)$  \\ \hline
\end{tabular}
\caption{Results show the mean and standard deviation of energy per token (J) and prefill energy (J) per inference for each LLM. These statistics are computed across the benchmark. }
\label{tab:resultsTokenCosts}
\end{table*}%
}

\newcommand{\tablePrefillModels}{%
\begin{table}[t!]
\centering
\scriptsize
\begin{tabular}{|lccccc|}
\hline
\multicolumn{1}{|c|}{\multirow{2}{*}{\textbf{Model}}}                                                                                                                                                                           & \multicolumn{5}{c|}{\textbf{Prefill Contribution (\%)}}                                                                                                                      \\ \cline{2-6} 
\multicolumn{1}{|c|}{}                                &  \multicolumn{1}{c|}{\textit{0-shot}} & \multicolumn{1}{c|}{\textit{2-shot}} & \multicolumn{1}{c|}{\textit{0-shot CoT}} & \multicolumn{1}{c|}{\textit{CU}} & \textit{CU-long} \\ \hline
\multicolumn{6}{|c|}{\textbf{6-7B Group}}                                                                                                                                                                                                                                                                                                                                                                                                                        \\ \hline
\multicolumn{1}{|l|}{CodeLlama-7B}                    & \multicolumn{1}{c|}{0.7}             & \multicolumn{1}{c|}{1.6}             & \multicolumn{1}{c|}{0.1}                 & \multicolumn{1}{c|}{84.4}        & 7.7              \\ \hline
\multicolumn{1}{|l|}{Qwen2.5-Coder-7B}                &  \multicolumn{1}{c|}{1.4}             & \multicolumn{1}{c|}{3.4}             & \multicolumn{1}{c|}{1.5}                 & \multicolumn{1}{c|}{74.9}        & 8.8              \\ \hline
\multicolumn{1}{|l|}{Deepseek-Coder-6.7B}             &  \multicolumn{1}{c|}{0.4}             & \multicolumn{1}{c|}{1.4}             & \multicolumn{1}{c|}{0.1}                 & \multicolumn{1}{c|}{73.4}        & 7.6              \\ \hline
\multicolumn{1}{|l|}{CodeGemma-7B}                    & \multicolumn{1}{c|}{1.2}             & \multicolumn{1}{c|}{2.8}             & \multicolumn{1}{c|}{0.9}                 & \multicolumn{1}{c|}{72.6}        & 22.8             \\ \hline
\multicolumn{1}{|l|}{CodeQwen1.5-7B}                  &  \multicolumn{1}{c|}{0.9}             & \multicolumn{1}{c|}{3.4}             & \multicolumn{1}{c|}{1.1}                 & \multicolumn{1}{c|}{79.9}        & 10.1             \\ \hline
\multicolumn{1}{|l|}{NextCoder-7B}                    &  \multicolumn{1}{c|}{0.8}             & \multicolumn{1}{c|}{1.2}             & \multicolumn{1}{c|}{0.5}                 & \multicolumn{1}{c|}{77.1}        & 20.5             \\ \hline
\multicolumn{6}{|c|}{\textbf{3-4B Group}}                                                                                                                                                                                                                                                                                                                                                                                                                        \\ \hline
\multicolumn{1}{|l|}{Phi3.5-4B}                       &  \multicolumn{1}{c|}{0.8}             & \multicolumn{1}{c|}{2.5}             & \multicolumn{1}{c|}{0.5}                 & \multicolumn{1}{c|}{72.1}        & 8.7              \\ \hline
\multicolumn{1}{|l|}{Phi4-4B}                         &  \multicolumn{1}{c|}{0.8}             & \multicolumn{1}{c|}{2.2}             & \multicolumn{1}{c|}{0.5}                 & \multicolumn{1}{c|}{67.3}        & 10.3             \\ \hline
\multicolumn{1}{|l|}{Qwen3-4B}                        & \multicolumn{1}{c|}{0.4}             & \multicolumn{1}{c|}{0.7}             & \multicolumn{1}{c|}{0.1}                 & \multicolumn{1}{c|}{68.3}        & 5.6              \\ \hline
\multicolumn{1}{|l|}{Qwen2.5-Coder-3B}                &  \multicolumn{1}{c|}{0.4}             & \multicolumn{1}{c|}{0.8}             & \multicolumn{1}{c|}{1.4}                 & \multicolumn{1}{c|}{64.9}        & 5.1              \\ \hline
\end{tabular}
\caption{\hl{Results report the mean contribution of the prefill phase to the total energy consumption during inference.} These values are computed across the benchmark. }
\label{tab:resultsPrefill}
\end{table}%
}

\newcommand{\tableTokenizers}{
\begin{table}[h!]
\scriptsize
\centering
\begin{tabular}{|l|c|c|c|c|c|}
\hline
\multicolumn{1}{|c|}{\textbf{Model}} & \textit{\textbf{0-shot}} & \textit{\textbf{2-shot}} & \textit{\textbf{0-shot CoT}} & \textit{\textbf{CU}} & \textit{\textbf{CU-long}} \\ \hline
\textbf{CodeLlama-7B}                & 163                      & 593                      & 169                          & 5555                 & 5503                      \\ \hline
\textbf{Qwen2.5-Coder-7B}            & 137                      & 504                      & 145                          & 4287                 & 4221                      \\ \hline
\textbf{Deepseek-Coder-6.7B}         & 162                      & 586                      & 169                          & 5542                 & 5491                      \\ \hline
\textbf{CodeGemma-7B}                & 155                      & 559                      & 161                          & 4888                 & 4812                      \\ \hline
\textbf{CodeQwen1.5-7B}              & 158                      & 582                      & 163                          & 5298                 & 5226                      \\ \hline
\textbf{NextCoder-7B}                & 136                      & 503                      & 146                          & 4278                 & 4204                      \\ \hline
\textbf{Phi3.5-4B}                   & 162                      & 593                      & 168                          & 5554                 & 5501                      \\ \hline
\textbf{Phi4-4B}                     & 134                      & 493                      & 143                          & 4144                 & 4078                      \\ \hline
\textbf{Qwen3-4B}                    & 138                      & 500                      & 144                          & 4244                 & 4184                      \\ \hline
\textbf{Qwen2.5-Coder-3B}            & 137                      & 505                      & 145                          & 4268                 & 4213                      \\ \hline
\end{tabular}
\caption{Mean number of input tokens for each model and each workload.}
\label{tab:tokenizers}
\end{table}
}

\newcommand{\tableEarlyStopping}{
\begin{table*}[t!]
\centering
\scriptsize
\begin{tabular}{|lccccccccc|}
\hline
\multicolumn{1}{|c|}{\multirow{2}{*}{}}      & \multicolumn{3}{c|}{\textbf{CodeLlama - 7B}}                                                                                   & \multicolumn{3}{c|}{\textbf{Deepseek - 6.7B}}                                                                                  & \multicolumn{3}{c|}{\textbf{Qwen3-4B}}                                                                    \\ \cline{2-10} 
\multicolumn{1}{|c|}{}                       & \multicolumn{1}{c|}{\textbf{Baseline}} & \multicolumn{1}{c|}{\textbf{Early Stopping}} & \multicolumn{1}{c|}{\textbf{$\Delta$}} & \multicolumn{1}{c|}{\textbf{Baseline}} & \multicolumn{1}{c|}{\textbf{Early Stopping}} & \multicolumn{1}{c|}{$\Delta$}          & \multicolumn{1}{c|}{\textbf{Baseline}} & \multicolumn{1}{c|}{\textbf{Early Stopping}} & $\Delta$          \\ \hline
\multicolumn{10}{|c|}{\textbf{max 300 tokens}}                                                                                                                                                                                                                                                                                                                                                                             \\ \hline
\multicolumn{1}{|l|}{\textbf{Output length}} & \multicolumn{1}{c|}{$205 (\pm 120)$}   & \multicolumn{1}{c|}{$114 (\pm 108)$}         & \multicolumn{1}{c|}{44\% $\textcolor{ForestGreen}{\downarrow}$} & \multicolumn{1}{c|}{$299 (\pm 1)$}     & \multicolumn{1}{c|}{$57 (\pm 52)$}           & \multicolumn{1}{c|}{81\% $\textcolor{ForestGreen}{\downarrow}$} & \multicolumn{1}{c|}{$299 (\pm 1)$}     & \multicolumn{1}{c|}{$125 (\pm 26)$}          & 59\% $\textcolor{ForestGreen}{\downarrow}$ \\ \hline
\multicolumn{1}{|l|}{\textbf{EpT}}           & \multicolumn{1}{c|}{$5.24 (\pm 0.09)$} & \multicolumn{1}{c|}{$8.01 (\pm 2.85)$}       & \multicolumn{1}{c|}{53\% $\textcolor{red}{\uparrow}$}   & \multicolumn{1}{c|}{$5.22 (\pm 0.04)$} & \multicolumn{1}{c|}{$8.42 (\pm 7.31)$}       & \multicolumn{1}{c|}{61\% $\textcolor{red}{\uparrow}$}   & \multicolumn{1}{c|}{$5.04 (\pm 0.03)$} & \multicolumn{1}{c|}{$6.55 (\pm 1.32)$}       & 30\% $\textcolor{red}{\uparrow}$   \\ \hline
\multicolumn{1}{|l|}{\textbf{Total}}         & \multicolumn{1}{c|}{$1075(\pm633)$}    & \multicolumn{1}{c|}{$1138(\pm1172)$}         & \multicolumn{1}{c|}{6\% $\textcolor{red}{\uparrow}$}    & \multicolumn{1}{c|}{$1562 (\pm 12)$}   & \multicolumn{1}{c|}{$483 (\pm 561)$}         & \multicolumn{1}{c|}{69\% $\textcolor{ForestGreen}{\downarrow}$} & \multicolumn{1}{c|}{$1508 (\pm 10)$}   & \multicolumn{1}{c|}{$840 (\pm 312)$}         & 44\% $\textcolor{ForestGreen}{\downarrow}$ \\ \hline
\multicolumn{1}{|l|}{\textbf{Accuracy}}      & \multicolumn{1}{c|}{34\%}              & \multicolumn{1}{c|}{32\%}                    & \multicolumn{1}{c|}{2\% $\textcolor{red}{\downarrow}$}  & \multicolumn{1}{c|}{61\%}              & \multicolumn{1}{c|}{63\%}                    & \multicolumn{1}{c|}{2\% $\textcolor{ForestGreen}{\uparrow}$}    & \multicolumn{1}{c|}{40\%}              & \multicolumn{1}{c|}{38\%}                    & 2\% $\textcolor{red}{\downarrow}$  \\ \hline
\multicolumn{10}{|c|}{\textbf{max 1000 tokens}}                                                                                                                                                                                                                                                                                                                                                                            \\ \hline
\multicolumn{1}{|l|}{\textbf{Output length}} & \multicolumn{1}{c|}{$625 (\pm 457)$}   & \multicolumn{1}{c|}{$127 (\pm123)$}          & \multicolumn{1}{c|}{80\% $\textcolor{ForestGreen}{\downarrow}$} & \multicolumn{1}{c|}{$981 (\pm99)$}     & \multicolumn{1}{c|}{$72 (\pm63)$}            & \multicolumn{1}{c|}{93\% $\textcolor{ForestGreen}{\downarrow}$} & \multicolumn{1}{c|}{$884(\pm215)$}     & \multicolumn{1}{c|}{$182(\pm118)$}           & 79\% $\textcolor{ForestGreen}{\downarrow}$ \\ \hline
\multicolumn{1}{|l|}{\textbf{EpT}}           & \multicolumn{1}{c|}{$5.33 (\pm 0.14)$} & \multicolumn{1}{c|}{$8.07 (\pm2.81)$}        & \multicolumn{1}{c|}{51\% $\textcolor{red}{\uparrow}$}   & \multicolumn{1}{c|}{$5.44 (\pm0.04)$}  & \multicolumn{1}{c|}{$8.26(\pm6.74)$}         & \multicolumn{1}{c|}{52\% $\textcolor{red}{\uparrow}$}   & \multicolumn{1}{c|}{$5.21(\pm0.11)$}   & \multicolumn{1}{c|}{$6.61 (\pm3.02)$}        & 27\% $\textcolor{red}{\uparrow}$   \\ \hline
\multicolumn{1}{|l|}{\textbf{Total}}         & \multicolumn{1}{c|}{$3379 (\pm 2484)$} & \multicolumn{1}{c|}{$1301 (\pm1385)$}        & \multicolumn{1}{c|}{62\% $\textcolor{ForestGreen}{\downarrow}$} & \multicolumn{1}{c|}{$5338 (\pm 550)$}  & \multicolumn{1}{c|}{$597(\pm654)$}           & \multicolumn{1}{c|}{89\% $\textcolor{ForestGreen}{\downarrow}$} & \multicolumn{1}{c|}{$4616(\pm1138)$}   & \multicolumn{1}{c|}{$1116 (\pm658)$}         & 76\% $\textcolor{ForestGreen}{\downarrow}$ \\ \hline
\multicolumn{1}{|l|}{\textbf{Accuracy}}      & \multicolumn{1}{c|}{33\%}              & \multicolumn{1}{c|}{32\%}                    & \multicolumn{1}{c|}{1\% $\textcolor{red}{\downarrow}$}  & \multicolumn{1}{c|}{59\%}              & \multicolumn{1}{c|}{63\%}                    & \multicolumn{1}{c|}{4\% $\textcolor{ForestGreen}{\uparrow}$}    & \multicolumn{1}{c|}{68\%}              & \multicolumn{1}{c|}{71\%}                    & 3\% $\textcolor{ForestGreen}{\uparrow}$    \\ \hline
\end{tabular}
\caption{Results for babbling models, where the baseline corresponds to the \texttt{0-shot} setting and early stopping denotes the application of the early stopping mechanism. Output length is in number of tokens. EpT is energy per token (in joules), along with total energy consumption per inference. $\Delta$ shows the difference between the baseline and early stopping. Red and green colors denote negative and positive impacts of early stopping, respectively.}
\label{tab:earlystopping}
\end{table*}
}

%% file: sample-base.bib
@inproceedings{methodology_babakol,
author = {Babakol, Timur and Liu, Yu David},
title = {Tensor-Aware Energy Accounting},
year = {2024},
isbn = {9798400702174},
publisher = {Association for Computing Machinery},
address = {New York, NY, USA},
url = {https://doi.org/10.1145/3597503.3639156},
booktitle = {Proceedings of the IEEE/ACM 46th International Conference on Software Engineering},
articleno = {93},
numpages = {12},
location = {Lisbon, Portugal},
series = {ICSE '24}
}

@ARTICLE{GenAISoftware,
  author={Ebert, Christof and Louridas, Panos},
  journal={IEEE Software}, 
  title={Generative AI for Software Practitioners}, 
  year={2023},
  volume={40},
  number={4},
  pages={30-38},
  doi={10.1109/MS.2023.3265877}}

@article{10.1145/3708525, author = {Shi, Jieke and Yang, Zhou and Lo, David}, title = {Efficient and Green Large Language Models for Software Engineering: Literature Review, Vision, and the Road Ahead}, year = {2025}, issue_date = {June 2025}, publisher = {Association for Computing Machinery}, address = {New York, NY, USA}, volume = {34}, number = {5}, issn = {1049-331X}, url = {https://doi.org/10.1145/3708525}, doi = {10.1145/3708525}, journal = {ACM Trans. Softw. Eng. Methodol.}, month = may, articleno = {137}, numpages = {22} }

@inbook{Basili,
author = {van Solingen (Revision), Rini and Basili (Original article, 1994 ed.), Vic and Caldiera (Original article, 1994 ed.), Gianluigi and Rombach (Original article, 1994 ed.), H. Dieter},
publisher = {John Wiley \& Sons, Ltd},
isbn = {9780471028956},
title = {Goal Question Metric (GQM) Approach},
booktitle = {Encyclopedia of Software Engineering},
chapter = {},
pages = {},
doi = {https://doi.org/10.1002/0471028959.sof142},
url = {https://onlinelibrary.wiley.com/doi/abs/10.1002/0471028959.sof142},
eprint = {https://onlinelibrary.wiley.com/doi/pdf/10.1002/0471028959.sof142},
year = {2002}
}

@misc{pynvml,
  author       = "{PyPI Contributors}",
  title        = "{pynvml: Python bindings for NVML}",
  year         = "2024",
  url          = "https://pypi.org/project/pynvml/",
  note         = "Accessed: 2025-10-23"
}

@misc{nvidia:2025,
      title={{NVIDIA Management Library (NVML)}}, 
      author={{NVIDIA Corporation}},
      year={2025},
      note={https://developer.nvidia.com/management-library-nvml. Last accessed October 22nd, 2025.}
}

@misc{githubCopilot, 
    author     = "GitHub", 
    title      = "GitHub Copilot", 
    year       = "2025", 
    url        = "https://copilot.github.com/",
    note       = "Accessed: 2025-10-23"
}

@misc{chatGPT, 
    author     = "OpenAI", 
    title      = "ChatGPT", 
    year       = "2025", 
    url        = "https://chat.openai.com/chat",
    note       = "Accessed: 2025-10-23"
}

@misc{babble, 
    author     = "Merriam-Webster", 
    title      = "babbling", 
    year       = "2026", 
    url        = "https://www.merriam-webster.com/dictionary/babbling",
    note       = "Accessed: 2026-01-22"
}

@misc{UTJupyterLab,
  author       = "{Univeristy of Twente}",
  title        = "{UT-JupyterLab Wiki}",
  year         = "2025",
  url          = "https://jupyter.wiki.utwente.nl",
  note         = "Accessed: 2025-10-23"
}

@misc{BigCodeLeaderboard, 
    author     = "BigCode", 
    title      = "", 
    year       = "2025", 
    url        = "https://huggingface.co/spaces/bigcode/bigcode-models-leaderboard",
    note       = "Accessed: 2025-10-23"
}

@misc{SemiAnalysis, 
    author     = "Dylan Patel and Afzal Ahmad", 
    title      = "The Inference Cost Of Search Disruption – Large Language Model Cost Analysis", 
    year       = "2023", 
    url        = "https://newsletter.semianalysis.com/p/the-inference-cost-of-search-disruption",
    note       = "Accessed: 2026-01-14"
}

@article{humanEval,
  author={Mark Chen and Jerry Tworek and Heewoo Jun and Qiming Yuan and Henrique Ponde de Oliveira Pinto and Jared Kaplan and Harri Edwards and others},
  title        = {Evaluating Large Language Models Trained on Code},
  journal      = {CoRR},
  volume       = {abs/2107.03374},
  year         = {2021},
}

@misc{longBench,
      title={LongBench v2: Towards Deeper Understanding and Reasoning on Realistic Long-context Multitasks}, 
      author={Yushi Bai and Shangqing Tu and Jiajie Zhang and Hao Peng and Xiaozhi Wang and Xin Lv and Shulin Cao and Jiazheng Xu and Lei Hou and Yuxiao Dong and Jie Tang and Juanzi Li},
      year={2025},
      eprint={2412.15204},
      archivePrefix={arXiv},
      primaryClass={cs.CL},
      url={https://arxiv.org/abs/2412.15204}, 
}

@inproceedings{fernandez-etal-2025-energy,
    title = "Energy Considerations of Large Language Model Inference and Efficiency Optimizations",
    author = "Fernandez, Jared  and
      Na, Clara  and
      Tiwari, Vashisth  and
      Bisk, Yonatan  and
      Luccioni, Sasha  and
      Strubell, Emma",
    editor = "Che, Wanxiang  and
      Nabende, Joyce  and
      Shutova, Ekaterina  and
      Pilehvar, Mohammad Taher",
    booktitle = "Proceedings of the 63rd Annual Meeting of the Association for Computational Linguistics (Volume 1: Long Papers)",
    month = jul,
    year = "2025",
    address = "Vienna, Austria",
    publisher = "Association for Computational Linguistics",
    url = "https://aclanthology.org/2025.acl-long.1563/",
    doi = "10.18653/v1/2025.acl-long.1563",
    pages = "32556--32569",
    ISBN = "979-8-89176-251-0"
}

@INPROCEEDINGS{alizadeh2025languagemodelssoftwaredevelopment,
  author={Alizadeh, Negar and Belchev, Boris and Saurabh, Nishant and Kelbert, Patricia and Castor, Fernando},
  booktitle={2025 IEEE/ACM 22nd International Conference on Mining Software Repositories (MSR)}, 
  title={Language Models in Software Development Tasks: An Experimental Analysis of Energy and Accuracy}, 
  year={2025},
  volume={},
  number={},
  pages={725-736},
  doi={10.1109/MSR66628.2025.00109}}

@ARTICLE{10549890,
  author={Argerich, Mauricio Fadel and Patiño-Martínez, Marta},
  journal={IEEE Access}, 
  title={Measuring and Improving the Energy Efficiency of Large Language Models Inference}, 
  year={2024},
  volume={12},
  number={},
  pages={80194-80207},
  doi={10.1109/ACCESS.2024.3409745}}

@article{DEVRIES20232191,
title = {The growing energy footprint of artificial intelligence},
journal = {Joule},
volume = {7},
number = {10},
pages = {2191-2194},
year = {2023},
issn = {2542-4351},
doi = {https://doi.org/10.1016/j.joule.2023.09.004},
url = {https://www.sciencedirect.com/science/article/pii/S2542435123003653},
author = {Alex {de Vries}}
}

@misc{maliakel2025,
      title={Investigating Energy Efficiency and Performance Trade-offs in LLM Inference Across Tasks and DVFS Settings}, 
      author={Paul Joe Maliakel and Shashikant Ilager and Ivona Brandic},
      year={2025},
      eprint={2501.08219},
      archivePrefix={arXiv},
      primaryClass={cs.LG},
      url={https://arxiv.org/abs/2501.08219}, 
}

@inproceedings{jenga,
author = {Zhang, Chen and Du, Kuntai and Liu, Shu and Kwon, Woosuk and Mo, Xiangxi and Wang, Yufeng and Liu, Xiaoxuan and You, Kaichao and Li, Zhuohan and Long, Mingsheng and Zhai, Jidong and Gonzalez, Joseph and Stoica, Ion},
title = {Jenga: Effective Memory Management for Serving LLM with Heterogeneity},
year = {2025},
isbn = {9798400718700},
publisher = {Association for Computing Machinery},
address = {New York, NY, USA},
url = {https://doi.org/10.1145/3731569.3764823},
doi = {10.1145/3731569.3764823},
booktitle = {Proceedings of the ACM SIGOPS 31st Symposium on Operating Systems Principles},
pages = {446–461},
numpages = {16},
location = {Lotte Hotel World, Seoul, Republic of Korea},
series = {SOSP '25}
}

@misc{jegham2025hungryai,
      title={How Hungry is AI? Benchmarking Energy, Water, and Carbon Footprint of LLM Inference}, 
      author={Nidhal Jegham and Marwan Abdelatti and Lassad Elmoubarki and Abdeltawab Hendawi},
      year={2025},
      eprint={2505.09598},
      archivePrefix={arXiv},
      primaryClass={cs.CY},
      url={https://arxiv.org/abs/2505.09598}, 
}

@inproceedings{wattsLuccioni,
author = {Luccioni, Sasha and Jernite, Yacine and Strubell, Emma},
title = {Power Hungry Processing: Watts Driving the Cost of AI Deployment?},
year = {2024},
isbn = {9798400704505},
publisher = {Association for Computing Machinery},
address = {New York, NY, USA},
url = {https://doi.org/10.1145/3630106.3658542},
doi = {10.1145/3630106.3658542},
booktitle = {Proceedings of the 2024 ACM Conference on Fairness, Accountability, and Transparency},
pages = {85–99},
numpages = {15},
location = {Rio de Janeiro, Brazil},
series = {FAccT '24}
}

@article{greenreviewLLMs,
author = {Shi, Jieke and Yang, Zhou and Lo, David},
title = {Efficient and Green Large Language Models for Software Engineering: Literature Review, Vision, and the Road Ahead},
year = {2025},
issue_date = {June 2025},
publisher = {Association for Computing Machinery},
address = {New York, NY, USA},
volume = {34},
number = {5},
issn = {1049-331X},
url = {https://doi.org/10.1145/3708525},
doi = {10.1145/3708525},
journal = {ACM Trans. Softw. Eng. Methodol.},
month = may,
articleno = {137},
numpages = {22}
}

@INPROCEEDINGS{ai-powered-power-hungry,
  author={Solovyeva, Lola and Weidmann, Sophie and Castor, Fernando},
  booktitle={2025 IEEE/ACM Second International Conference on AI Foundation Models and Software Engineering (Forge)}, 
  title={AI-Powered, But Power-Hungry? Energy Efficiency of LLM-Generated Code}, 
  year={2025},
  volume={},
  number={},
  pages={49-60},
  doi={10.1109/Forge66646.2025.00012}}

@inproceedings{humanevalPlus,
 author = {Liu, Jiawei and Xia, Chunqiu Steven and Wang, Yuyao and ZHANG, LINGMING},
 booktitle = {Advances in Neural Information Processing Systems},
 editor = {A. Oh and T. Naumann and A. Globerson and K. Saenko and M. Hardt and S. Levine},
 pages = {21558--21572},
 publisher = {Curran Associates, Inc.},
 title = {Is Your Code Generated by ChatGPT Really Correct? Rigorous Evaluation of Large Language Models for Code Generation},
 url = {https://proceedings.neurips.cc/paper_files/paper/2023/file/43e9d647ccd3e4b7b5baab53f0368686-Paper-Conference.pdf},
 volume = {36},
 year = {2023}
}

@article{hallucinationsCodeGen, author = {Zhang, Ziyao and Wang, Chong and Wang, Yanlin and Shi, Ensheng and Ma, Yuchi and Zhong, Wanjun and Chen, Jiachi and Mao, Mingzhi and Zheng, Zibin}, title = {LLM Hallucinations in Practical Code Generation: Phenomena, Mechanism, and Mitigation}, year = {2025}, issue_date = {July 2025}, publisher = {Association for Computing Machinery}, address = {New York, NY, USA}, volume = {2}, number = {ISSTA}, url = {https://doi.org/10.1145/3728894}, doi = {10.1145/3728894}, journal = {Proc. ACM Softw. Eng.}, month = jun, articleno = {ISSTA022}, numpages = {23} }

@INPROCEEDINGS{codeQualityLLM,
  author={Jamil, Mohammad Talal and Abid, Shamsa and Shamail, Shafay},
  booktitle={2025 IEEE/ACM 22nd International Conference on Mining Software Repositories (MSR)}, 
  title={Can LLMs Generate Higher Quality Code Than Humans? An Empirical Study}, 
  year={2025},
  volume={},
  number={},
  pages={478-489},
  doi={10.1109/MSR66628.2025.00081}}

@misc{apsan2025,
      title={Generating Energy-Efficient Code via Large-Language Models -- Where are we now?}, 
      author={Radu Apsan and Vincenzo Stoico and Michel Albonico and Rudra Dhar and Karthik Vaidhyanathan and Ivano Malavolta},
      year={2025},
      eprint={2509.10099},
      archivePrefix={arXiv},
      primaryClass={cs.SE},
      url={https://arxiv.org/abs/2509.10099}, 
}

@INPROCEEDINGS{11113611,
  author={Fan, Haoyang and Lin, Yi-Chien and Prasanna, Viktor},
  booktitle={2025 IEEE 36th International Conference on Application-specific Systems, Architectures and Processors (ASAP)}, 
  title={ELLIE: Energy-Efficient LLM Inference at the Edge Via Prefill-Decode Splitting}, 
  year={2025},
  volume={},
  number={},
  pages={139-146},
  doi={10.1109/ASAP65064.2025.00031}}

@article{10.1145/3757892.3757900, author = {Niu, Chenxu and Zhang, Wei and Zhao, Yongjian and Chen, Yong}, title = {Energy Efficient or Exhaustive? Benchmarking Power Consumption of LLM Inference Engines}, year = {2025}, issue_date = {July 2025}, publisher = {Association for Computing Machinery}, address = {New York, NY, USA}, volume = {5}, number = {2}, url = {https://doi.org/10.1145/3757892.3757900}, doi = {10.1145/3757892.3757900}, journal = {SIGENERGY Energy Inform. Rev.}, month = aug, pages = {56–62}, numpages = {7} }

@misc{stojkovic2024,
      title={Towards Greener LLMs: Bringing Energy-Efficiency to the Forefront of LLM Inference}, 
      author={Jovan Stojkovic and Esha Choukse and Chaojie Zhang and Inigo Goiri and Josep Torrellas},
      year={2024},
      eprint={2403.20306},
      archivePrefix={arXiv},
      primaryClass={cs.AI},
      url={https://arxiv.org/abs/2403.20306}, 
}

@inproceedings{10.1145/3643795.3648379,
author = {Rasnayaka, Sanka and Wang, Guanlin and Shariffdeen, Ridwan and Iyer, Ganesh Neelakanta},
title = {An Empirical Study on Usage and Perceptions of LLMs in a Software Engineering Project},
year = {2024},
isbn = {9798400705793},
publisher = {Association for Computing Machinery},
address = {New York, NY, USA},
url = {https://doi.org/10.1145/3643795.3648379},
doi = {10.1145/3643795.3648379},
booktitle = {Proceedings of the 1st International Workshop on Large Language Models for Code},
pages = {111–118},
numpages = {8},
location = {Lisbon, Portugal},
series = {LLM4Code '24}
}

@INPROCEEDINGS{10628428,
  author={Jahić, Jasmin and Sami, Ashkan},
  booktitle={2024 IEEE 21st International Conference on Software Architecture Companion (ICSA-C)}, 
  title={State of Practice: LLMs in Software Engineering and Software Architecture}, 
  year={2024},
  volume={},
  number={},
  pages={311-318},
  doi={10.1109/ICSA-C63560.2024.00059}}

@inproceedings{10.1145/3696630.3730563,
author = {Zakharov, Ilya and Koshchenko, Ekaterina and Sergeyuk, Agnia},
title = {AI in Software Engineering: Perceived Roles and Their Impact on Adoption},
year = {2025},
isbn = {9798400712760},
publisher = {Association for Computing Machinery},
address = {New York, NY, USA},
url = {https://doi.org/10.1145/3696630.3730563},
doi = {10.1145/3696630.3730563},
booktitle = {Proceedings of the 33rd ACM International Conference on the Foundations of Software Engineering},
pages = {1305–1309},
numpages = {5},
location = {Clarion Hotel Trondheim, Trondheim, Norway},
series = {FSE Companion '25}
}

@inproceedings{10.1145/3639477.3643648,
author = {Davila, Nicole and Wiese, Igor and Steinmacher, Igor and Lucio da Silva, Lucas and Kawamoto, Andre and Favaro, Gilson Jose Peres and Nunes, Ingrid},
title = {An Industry Case Study on Adoption of AI-based Programming Assistants},
year = {2024},
isbn = {9798400705014},
publisher = {Association for Computing Machinery},
address = {New York, NY, USA},
url = {https://doi.org/10.1145/3639477.3643648},
doi = {10.1145/3639477.3643648},
booktitle = {Proceedings of the 46th International Conference on Software Engineering: Software Engineering in Practice},
pages = {92–102},
numpages = {11},
location = {Lisbon, Portugal},
series = {ICSE-SEIP '24}
}

@inproceedings{10.5555/3691938.3691945, author = {Agrawal, Amey and Kedia, Nitin and Panwar, Ashish and Mohan, Jayashree and Kwatra, Nipun and Gulavani, Bhargav S. and Tumanov, Alexey and Ramjee, Ramachandran}, title = {Taming throughput-latency tradeoff in LLM inference with sarathi-serve}, year = {2024}, isbn = {978-1-939133-40-3}, publisher = {USENIX Association}, address = {USA}, booktitle = {Proceedings of the 18th USENIX Conference on Operating Systems Design and Implementation}, articleno = {7}, numpages = {18}, location = {Santa Clara, CA, USA}, series = {OSDI'24} }

@article{Verdecchia2023ASR,
  title={A systematic review of Green AI},
  author={Roberto Verdecchia and June Sallou and Lu{\'i}s Cruz},
  journal={Wiley Interdisciplinary Reviews: Data Mining and Knowledge Discovery},
  year={2023},
  volume={13},
  url={https://api.semanticscholar.org/CorpusID:256274885}
}

@INPROCEEDINGS{splitwise,
  author={Patel, Pratyush and Choukse, Esha and Zhang, Chaojie and Shah, Aashaka and Goiri, Íñigo and Maleki, Saeed and Bianchini, Ricardo},
  booktitle={2024 ACM/IEEE 51st Annual International Symposium on Computer Architecture (ISCA)}, 
  title={Splitwise: Efficient Generative LLM Inference Using Phase Splitting}, 
  year={2024},
  volume={},
  number={},
  pages={118-132},
  doi={10.1109/ISCA59077.2024.00019}}

@misc{spad,
      title={SPAD: Specialized Prefill and Decode Hardware for Disaggregated LLM Inference}, 
      author={Hengrui Zhang and Pratyush Patel and August Ning and David Wentzlaff},
      year={2025},
      eprint={2510.08544},
      archivePrefix={arXiv},
      primaryClass={cs.AR},
      url={https://arxiv.org/abs/2510.08544}, 
}

@inproceedings{distserve,
author = {Zhong, Yinmin and Liu, Shengyu and Chen, Junda and Hu, Jianbo and Zhu, Yibo and Liu, Xuanzhe and Jin, Xin and Zhang, Hao},
title = {DistServe: disaggregating prefill and decoding for goodput-optimized large language model serving},
year = {2024},
isbn = {978-1-939133-40-3},
publisher = {USENIX Association},
address = {USA},
booktitle = {Proceedings of the 18th USENIX Conference on Operating Systems Design and Implementation},
articleno = {11},
numpages = {18},
location = {Santa Clara, CA, USA},
series = {OSDI'24}
}

@misc{phaseLevelEnergyAnalysis,
  title        = {Anonymized GitHub Repository: PhaseLevelEnergyAnalysisLLMInference},
author={Anonymous},
  year         = {2026},
  publisher    = {Anonymous GitHub},
  url          = {https://anonymous.4open.science/r/PhaseLevelAnalysisLLMInference-4E96},
  note         = {Accessed: Jan. 25, 2026}
}

@ARTICLE{11305123,
author={Rajput, Saurabhsingh and Saad, Mootez and Sharma, Tushar},
journal={ IEEE Software },
title={{ Tu(r)ning AI Green: Exploring Energy Efficiency Cascading with Orthogonal Optimizations }},
year={5555},
volume={},
number={01},
ISSN={1937-4194},
pages={1-7},
doi={10.1109/MS.2025.3645090},
url = {https://doi.ieeecomputersociety.org/10.1109/MS.2025.3645090},
publisher={IEEE Computer Society},
address={Los Alamitos, CA, USA},
month=dec}

@misc{saad2025senaisoftwareengineeringnative,
      title={SENAI: Towards Software Engineering Native Generative Artificial Intelligence}, 
      author={Mootez Saad and José Antonio Hernández López and Boqi Chen and Neil Ernst and Dániel Varró and Tushar Sharma},
      year={2025},
      eprint={2503.15282},
      archivePrefix={arXiv},
      primaryClass={cs.SE},
      url={https://arxiv.org/abs/2503.15282}, 
}

@misc{mehditabar2025smartcostlybenchmarkingllms,
      title={Smart but Costly? Benchmarking LLMs on Functional Accuracy and Energy Efficiency}, 
      author={Mohammadjavad Mehditabar and Saurabhsingh Rajput and Antonio Mastropaolo and Tushar Sharma},
      year={2025},
      eprint={2511.07698},
      archivePrefix={arXiv},
      primaryClass={cs.SE},
      url={https://arxiv.org/abs/2511.07698}, 
}
